\documentclass[12pt]{article}
\usepackage{setspace}
\usepackage{float}
\usepackage{amsmath}
\usepackage{graphicx}
\usepackage{epstopdf}
\usepackage{hyperref}

\begin{document}
\title{Nonparametric CUSUM Charts for Circular Data with Applications in Health Science and Astrophysics}
\author{F. Lombard\\ Department of Statistics, University of Johannesburg \\ fredl@uj.ac.za\\
\\
        Douglas M. Hawkins\\
                      Scottsdale Scientific LLC, Scottsdale, AZ
                       \\ dhawkins@umn.edu\\
\\
        Cornelis J. Potgieter\\
                   Department of Statistical Science\\Southern Methodist University, Dallas, TX\\ \& Department of Statistics, University of Johannesburg \\ cpotgieter@smu.edu
}
\date{}
\maketitle

\begin{abstract}
	This paper develops non-parametric rotation invariant CUSUMs suited to the
	detection of changes in the mean direction as well as changes in the
	concentration parameter of circular data. The properties of the CUSUMs are
	illustrated by theoretical calculations, Monte Carlo simulation and
	application to sequentially observed angular data from health science and astrophysics.
	
\end{abstract}




\doublespacing
\section{Introduction}

\label{Introduction}

Sequential CUSUM methods for detecting parameter changes in distributions on
the real line is a well developed field with an extensive literature. The same
cannot be said about CUSUM methods to detect changes of location in
non-Euclidean spaces such as the circle. Distributions on the circle generate
data which cannot generally be treated in the same manner as linear data - see
Fisher (1993, Chapter 1 and Section 3.1), Mardia and Jupp (2000, Chapter 1)
and Jammalamadaka and SenGupta (2001, Section 1.2.2). One impediment to the
application of linear CUSUM methods is the fact that a circle has no well
separated beginning and end. Whichever point is selected as the beginning
point, the distance between it and the endpoint is zero. A family of
distributions with a fixed arc on the circle as support could in principle be
treated as if the sample space were a finite fixed interval on the real line.
However, the options involved in formulating a changepoint model would then be
severely curtailed: a model involving shifts of arbitrary size in the location
of the distribution would be out of the question. The distributions from which
the data in our applications in Section \ref{Examples} arise encompass the
full circle and are therefore not amenable to analysis by linear CUSUM methods.

Lombard, Hawkins and Potgieter (2018) reviewed the current state of change
detection procedures for circular data. They also constructed distribution
free CUSUMs for circular data in which the numerical value of an in-control
mean direction is specified, the objective being to detect a change in mean
direction away from this value. The situation is analogous to that in which
the well known Page (1954) CUSUM is applied, namely detection of a change away
from a specified numerical value of the mean of a distribution on the real
line. However, in the examples treated in Section \ref{Examples} of the
present paper, no in-control circular mean value is specified and the
objective is to detect a change away from the unknown current circular mean
value, whatever it may be. Such a CUSUM, unlike that proposed by Lombard,
Hawkins and Potgieter (2018), must be rotation invariant because the outcome
of the analysis should not depend upon which point on the circle is chosen as
the origin of angular measurement.

The main contribution of the present paper is the construction of such
invariant CUSUMs for circular data. The CUSUMs we construct are non-parametric
in the sense that their form is not dependent upon an underlying
parametrically specified distribution. The in-control properties of the CUSUMs
are shown in a Monte Carlo study to be quite robust over a wide class of
circular distributions, which makes them near distribution free over this
class. As far as we are aware, no CUSUMs of this nature for circular data have
to date been treated in the statistical literature.

Section \ref{Direction cusum} of the paper focuses on mean direction. We
provide justifications for\ the form of our CUSUM and discuss some
computational details. In Section \ref{In control properties} we elaborate on
its in-control and out-of control properties. The results of an extensive
Monte Carlo study are also reported. In Section \ref{Concentration change} we
briefly consider a CUSUM for detecting concentration changes. Section
\ref{Examples} demonstrates the application of the CUSUMs to two sets of data
and Section \ref{Summary} summarizes our results.

\section{Detecting direction change}

\label{Direction cusum}

\subsection{Derivation of the CUSUM statistic}

\label{Motivation}

Initially the data $X_{1},X_{2},\ldots$ come from a non-uniform and unimodal
continuous distribution $F$ with unknown mean direction $\nu=\nu_{0}$ on the
circle $[-\pi,\pi)$. This defines the in-control state. (Since mean direction
is a vacuous concept in a uniform distribution, the latter is excluded from
consideration. The CUSUM of Lombard and Maxwell (2012), which is rotation
invariant, can be used to detect a change from a uniform to a non-uniform
distribution.) We estimate $\nu$ by
\begin{equation}
\hat{\nu}_{n}=\text{atan2}(S_{n},C_{n}) \label{mean direction}%
\end{equation}
where for $n=2,3,\ldots$,%

\begin{equation}
C_{n}=\sum_{j=1}^{n} \cos X_{j},\qquad S_{n}=\sum_{j=1}^{n}\sin X_{j}, \label{C S defns}%
\end{equation}
and atan2 denotes the four-quadrant inverse tangent function%

\[
\text{atan2}(x,y)=\left\{
\begin{array}
[c]{ll}%
\tan^{-1}(x/y) & \mathrm{if}\ y>0\\
\tan^{-1}(x/y)+\pi \mathrm{sign}(x) & \mathrm{if}\ y<0\\
(\pi/2)\mathrm{sign}(x) & \mathrm{if}\ y=0,x\neq0\\
0 & \mathrm{if}\ y=x=0,
\end{array}
\right. %
\]
the symbol $\tan^{-1}$ denoting the usual inverse tangent function with range
restricted to $(-\pi/2,\pi/2)$. This non-parametric estimator is, in fact,
also the maximum likelihood estimator of mean direction in a von Mises
distribution, which is arguably the best known among circular distributions.
The von Mises distribution with mean direction $\nu$ and concentration
$\kappa$, has density function%

\[
f(x)=\frac{1}{2\pi I_{0}(\kappa)}\exp[\kappa\cos(x-\nu)],\quad -\pi\leq x<\pi,
\]
where $I_{0}$ denotes the modified Bessel function of the
first kind of order zero. The log-likelihood ratio based on observations
$X_{1}+\delta,\ldots,X_{n}+\delta$ is, apart from a factor not depending upon
$\delta$, given by%
\[
l(\delta)=\cos(X_{n}-\delta-\nu)
\]
and a locally most powerful test of the hypothesis $H_{0}:\delta=0$ is
therefore based on the derivative%
\[
\left.  \frac{dl(\delta)}{d\delta}\right|  _{\delta=0}=\sin(X_{n}-\nu).
\]
Replacing $\nu$ by $\hat{\nu}_{n-1}$ leads to consideration of a CUSUM based
on the statistic
\begin{equation}
V_{n}=\sin(X_{n}-\hat{\nu}_{n-1}). \label{V_n}%
\end{equation}
Despite the fact that $V_{n}$ originates from the von Mises distribution, it
has at least two purely non-parametric origins that do not depend upon any
assumption involving the type of the underlying distribution.

The first of these follows upon expanding the sine function and using the
trigonometric relations%
\[
\sin(\hat{\nu}_{n-1})=S_{n-1}/R_{n-1},\qquad \cos(\hat{\nu}_{n-1})=C_{n-1}%
/R_{n-1},
\]
wherein
\begin{equation}
R_{n}^{2}=C_{n}^{2}+S_{n}^{2}. \label{R^2}%
\end{equation}
This gives
\begin{equation}
V_{n}=(C_{n-1}/R_{n-1})\sin X_{n}-(S_{n-1}/R_{n-1})\cos X_{n},
\label{V_n second form}%
\end{equation}
which is the (signed) area of the parallelogram spanned by the unit length
vectors $(C_{n-1},S_{n-1})/R_{n}$ and $(\sin X_{n},\cos X_{n})$. The former
of these vectors points in the mean direction of the data $X_{1}%
,\ldots,X_{n-1}$ while the latter vector points in the direction of the new
observation $X_{n}$.and the greater the angular distance between the two
directions is, the larger will be the area of the parallelogram. Thus, if a
change in mean direction $\nu$ occurs at index $n$, we can expect a succession
of positive or negative values $V_{n},\ n>\tau$.

A second non-parametric argument leading to consideration of $V_{n}$ comes
from considering the change $\hat{\nu}_{n}-\hat{\nu}_{n-1}$ in the estimate of
$\nu$ effected by a change in mean direction from $\nu$ to $\nu+\delta$
occurring at index $n$. We have%
\begin{align*}
\hat{\nu}_{n}  &  =\text{atan2}\left[S_{n-1}+\sin (X_{n}+\delta),C_{n-1}%
+\cos (X_{n}+\delta)\right] \\
& \\
&  =\text{atan2}(S_{n-1}/n+\delta_{1,n},C_{n-1}/n+\delta_{2,n})
\end{align*}
where%
\begin{align*}
n\delta_{1,n}  &  =\sin (X_{n}+\delta)=\sin X_{n}+O(\delta)_{,}\\
& \\
n\delta_{2,n}  &  =\cos (X_{n}+\delta)=\cos X_{n}+O(\delta)_{.}%
\end{align*}
Since both $S_{n-1}/n$ and $C_{n-1}/n$ converge as $n\rightarrow\infty$, and
both $\delta_{1,n}$ and $\delta_{2,n}$ tend to zero, we can make a Taylor
expansion around $(S_{n-1}/n,C_{n-1}/n)$. This gives%
\begin{align*}
R_{n-1}(\hat{\nu}_{n}-\hat{\nu}_{n-1})  &  =n\delta_{1,n}\frac{C_{n-1}%
}{R_{n-1}}-n\delta_{2,n}\frac{S_{n-1}}{R_{n-1}}+O(n^{-1})\\
& \\
&  =\frac{C_{n-1}}{R_{n-1}}\sin X_{n}-\frac{S_{n-1}}{R_{n-1}}\cos
 X_{n}+O(\delta)+O(n^{-1})\\
& \\
&  =V_{n}+O(\delta)+O(n^{-1}),%
\end{align*}
which shows again the relevance of $V_{n}$ for detecting changes in mean direction.

The most important property of $V_{n}$ as far as motivation for the present
paper is concerned is its rotation invariance: its numerical values are
unaffected if \textit{all} the data are rotated through the same fixed, but
unknown, angle. Thus, a CUSUM based on $V_{n}$ will be applicable in
situations where no in-control direction is specified and the objective is
merely to detect deviations from this arbitrary in-control direction. Both
examples treated in Section \ref{Examples} of the paper are of this nature.
This contrasts with the distribution free CUSUMs in Lombard, Hawkins and
Potgieter (2017), which require a specified numerical value of the in-control
mean direction.

\subsection{Construction of the CUSUM}

\label{Construction}

When the process is in control, that is, when $X_{1},X_{2},\ldots$ are
independently and identically distributed (but with unknown mean direction),
then%
\begin{equation}
\xi_{n}:=(V_{n}-\mathrm{E}_{n-1}\left[  V_{n}\right]  )/\sqrt{\mathrm{Var}%
	_{n-1}\left[  V_{n}\right]  },\ n\geq2, \label{xi raw}%
\end{equation}
is a martingale difference sequence with conditional variance $1$. Here and
elsewhere, E$_{n-1}[\cdot]$ and Var$_{n-1}[\cdot]$ denote expected value and
variance computed conditionally upon $X_{1},\ldots,X_{n-1}$. Using standard
martingale central limit theory, we can show that cumulative sums of the
$\xi_{n}$ will be asymptotically normally distributed regardless of the type
of underlying distribution - see, e.g. Helland (1982, Theorem 3.2).
Furthermore, if $\nu=\nu_{0}$ changes by an amount $\delta$ to $\nu=\nu
_{0}+\delta$ at observation $X_{\tau+1}$ ($\tau$ being the last in-control
observation) then by either of the two arguments following (\ref{V_n}), we can
expect E$_{\tau}\left[  \xi_{\tau+1}\right]  $ to be non-zero. Thus, a
standard two-sided normal CUSUM for data on the real line, applied to the
$\xi_{n}$ sequence, could be expected to be effective in detecting a change
away from the initial direction. Furthermore, the in-control behaviour should
be \textit{quantitatively} similar to that of a standard normal CUSUM.

The conditional mean and variance in (\ref{xi raw}) depend on the first two
moments of $\sin X$ and $\cos X$, which are unknown parameters. Accordingly,
given observations $X_{1},\ldots,X_{n}$, we estimate the conditional mean and
variance non-parametrically by%
\[
\hat{\mathrm{{E}}}_{n-1}\left[  V_{n}\right]  =\frac{1}{n-1}\sum\nolimits_{i=1}%
^{n-1}\sin(X_{i}-\hat{\nu}_{n-1})=0
\]
and%
\begin{equation}
\widehat{\mathrm{Var}}_{n-1}\left[  V_{n}\right]  =\frac{1}{n-1}\sum
\nolimits_{i=1}^{n-1}\sin^{2}(X_{i}-\hat{\nu}_{n-1}):=B_{n-1}^{2}.
\label{var_hat_V_1}%
\end{equation}
Then a computable CUSUM is obtained upon replacing $\xi_{n}$ in (\ref{xi raw})
by
\begin{equation}
\hat{\xi}_{n}=V_{n}/B_{n-1}. \label{xi direction cusum}%
\end{equation}
The CUSUM is started at observation $m+1$ by setting $D_{i}^{\pm}=0$ for
$i=1,\ldots,m$ and
\begin{align}
D_{m+n}^{+}  &  =\max\{0,D_{m+n-1}+\hat{\xi}_{m+n}-\zeta\}\nonumber\\
& \label{cusums}\\
D_{m+n}^{-}  &  =\min\{0,D_{m+n-1}+\hat{\xi}_{m+n}+\zeta\}\nonumber
\end{align}
for $n\geq1$, where $\zeta$ is the reference value. The run length, $N$, is
the first index $n$ at which either $D_{m+n}^{+}\geq h$ or $D_{m+n}^{-}\leq
-h$, where $h$ is a control limit. The control limit is chosen to produce a
specified in-control average run length (ARL), which we denote throughout by
$ARL_{0}$. The first $m$ observations serve to make an initial estimate of the
population moments \textit{after which the estimates are updated with the
	arrival of each new observation}. Since the the random variables $\sin\ X$ and
$\cos\ X$ are bounded, convergence of sample moments to population moments
would be quite rapid so that a relatively small number $m$ of observations
should suffice to initialize the CUSUM.

\subsection{Implementation}

\label{Implementation}

Implementation of the CUSUM scheme requires an efficient method of updating
the summand $\hat{\xi}_{n-1}$ upon arrival of a new observation $X_{n}$. For
this, set%
\[
s_{n}=\sin X_{n},\quad c_{n}=\cos X_{n}%
\]
and%
\[
C_{n}^{(2)}=\sum_{j=1}^{n} c_{j}^{2},\quad 
S_{n}^{(2)}=\sum_{j=1}^{n} s_{j}^{2},\quad
A_{n}^{(2)}=\sum_{j=1}^{n} s_{j}c_{j}%
\]
and observe that
\begin{equation}
(n-1)B_{n-1}^{2}=\frac{C_{n-1}^{2}}{R_{n-1}^{2}}S_{n-1}^{(2)}+\frac
{S_{n-1}^{2}}{R_{n-1}^{2}}C_{n-1}^{(2)}-2\frac{C_{n-1}S_{n-1}}{R_{n-1}^{2}%
}A_{n-1}^{(2)}. \label{B_n-1 expression}%
\end{equation}
In particular, we see that the $R_{n-1}$ factors in $V_{n}$ and $B_{n-1}$
cancel, whence%
\begin{equation}
\hat{\xi}_{n}=\frac{V_{n}^{\ast}}{B_{n-1}^{\ast}}:=\frac{C_{n-1}\sin
	X_{n}-S_{n-1}\cos\ X_{n}}{\sqrt{\left(  C_{n-1}^{2}S_{n-1}^{(2)}+S_{n-1}%
		^{2}C_{n-1}^{(2)}-2C_{n-1}S_{n-1}A_{n-1}^{(2)}\right)  /(n-1)}}_{.}
\label{xi modified}%
\end{equation}
Next, note the simple recursions
\[
S_{n-1}=S_{n-2}+s_{n-1},\quad C_{n-1}=C_{n-2}+c_{n-1},
\]%
\[
S_{n-1}^{(2)}=S_{n-2}^{(2)}+s_{n-1}^{2},\quad C_{n-1}^{(2)}=C_{n-2}^{(2)}%
+c_{n-1}^{2}%
\]
and%
\[
A_{n-1}^{(2)}=A_{n-2}^{(2)}+s_{n-1}c_{n-1}.
\]
To compute $V_{n}^{\ast}$ in (\ref{xi modified}) given $S_{n-2},C_{n-2}%
,c_{n-1},c_{n},s_{n-1}$ and $s_{n}$, use the first of these recursions. To
compute $B_{n-1}$, given $S_{n-1},C_{n-1},S_{n-1}^{(2)},C_{n-1}^{(2)}%
,A_{n-1}^{(2)},c_{n-1},$ and $s_{n-1}$, use (\ref{B_n-1 expression}).

A rational basis for specifying a reference value $\zeta$ is also required.
This aspect of the CUSUM design is considered in Section
\ref{Choice of ref const} of the paper.

\section{In-control properties}

\label{In control properties}

While the proposed CUSUM is not distribution free, the asymptotic in-control
normality of CUSUMs of $\hat{\xi}_{n}$ suggests that it may be nearly so.
Then, use of standard normal distribution CUSUM control limits should lead to
an in-control ARL sufficiently close to the nominal value to make the CUSUMs
of practical use. The requisite control limit $h$ can be obtained from the
widely available software packages of Hawkins, Olwell and Wang, (2016) or
Knoth (2016). To check this expectation we estimated by Monte Carlo simulation
the in-control ARL over a range of unimodal symmetric and asymmetric
distributions on the circle. Among the multitude of possible distributions,
the class of wrapped stable and Student $t$ distributions, together with their
skew versions, represent a wide range of unimodal distribution shapes on the
circle. Simulated data from these distributions are easily obtained by
generating random numbers $Y$ from the distribution on the real line and then
wrapping these around the circle by the simple transformation
$Y(\operatorname{mod}\ 2\pi)$. Algorithms for generating the random numbers
$Y$ are given in Nolan (2015) and in Azzalini and Capitanio (2003). The
algorithms were implemented in Matlab and the relevant programs are included
in the supplementary material to this paper.

Some simulations were also run on data from other types of distribution which
are defined directly on the circle and not obtained by wrapping. Specifically,
we used the sine-skewed\ distributions developed Umbach and Jammalamadaka
(2011) and by Abe and Pewsey (2011). In contrast to the wrapped stable and
Student $t$ distributions, the densities of these distributions have closed
form expressions, which facilitates model fitting and parameter estimation.
The various unimodal distribution shapes available in these classes of
distributions are quite similar to those in the class of wrapped
distributions. Since the behaviour of a non-parametric CUSUM depends more on
the general shape of the underlying distribution than on the specific
parameter values producing that shape, it comes as no surprise that the
in-control behaviour of the CUSUMs proposed here is quite similar in the two
classes (wrapped and directly constructed) of distributions. Since wrapped
distributions are widely known and understood, we frame our discussion in the
context of these distributions. Some simulation results for data from the
sine-skewed distributions are included in the supplementary material to this
paper. In the discussion that follows, $S_{\alpha},\ 0<\alpha\leq2$, denotes a
stable distribution with index $\alpha$ and $t_{n},\ n\geq1$ denotes a Student
$t$-distribution with $n$ degrees of freedom.

In assessing the performance of the direction CUSUM under various symmetric
in-control and out-of-control distributions, we standardize the observations
to a common measure of concentration. The concentration parameter $\kappa$ of
the von Mises($\nu,\kappa$) distribution satisfies the relation
\begin{equation}
\kappa=A^{-1}(\text{E}[\cos(X-\nu)]) \label{kappa defn}%
\end{equation}
where $A(\kappa)=I_{1}(\kappa)/I_{0}(\kappa)$ and $I_{1}$ denotes the modified
Bessel function of the first kind of order $1$. In view of the status of the
von Mises distribution among circular distributions, which is much like that
of the normal distribution among distributions on the real line, we use in
this paper $\kappa$ in (\ref{kappa defn}) as a measure of the concentration of
a unimodal circular distribution with mean direction $\nu$. Thus, given
$\kappa$ and the density function of $Y$, the scale parameter $\sigma$ is
chosen to make the distribution of the wrapped random variable%
\[
X=(\sigma Y)_{w}:=\sigma Y (\operatorname{mod} 2\pi)
\]
satisfy (\ref{kappa defn}).

For instance, suppose $Y$ has an $S_{\alpha}$ distribution with characteristic
function
\[
\phi(t;\alpha)=\text{E}[\cos tY]=\exp(-|t|^{\alpha}).
\]
Then (Jammallamadaka and SenGupta, 2001, Proposition 2.1),
\[
\text{E}[\cos (\sigma Y)_{w}]=\phi(\sigma;\alpha)=\exp(-\sigma^{\alpha})
\]
so that
\begin{equation}
\sigma=(-\log(\ A(\kappa)))^{1/\alpha}. \label{kappa to stable sig}%
\end{equation}
As another example, a Student $t$-distribution with $\alpha$ degrees of
freedom has characteristic function
\[
\phi(t;\alpha)=\frac{K_{\alpha/2}(\sqrt{\alpha}t)(\sqrt{\alpha}t)^{\alpha/2}%
}{2^{\alpha/2-1}\Gamma(\frac{\alpha}{2})}%
\]
where $K_{\alpha/2}$ denotes the modified Bessel function of the second kind
order $\alpha/2$ and $\Gamma$ denotes the gamma function. Thus, in this case,
\[
\text{E}[\cos (\sigma Y)_{w}]=\phi(\sigma;\alpha)=\frac{K_{\alpha/2}%
	(\sqrt{\alpha}\sigma)(\sqrt{\alpha}\sigma)^{\alpha/2}}{2^{\alpha/2-1}%
	\Gamma(\frac{\alpha}{2})}_{,}%
\]
and $\sigma$ is the solution to the equation
\begin{equation}
K_{\alpha/2}(\sqrt{\alpha}\sigma)(\sqrt{\alpha}\sigma)^{\alpha/2}%
=2^{\alpha/2-1}\Gamma(\frac{\alpha}{2})A(\kappa). \label{kappa to student sig}%
\end{equation}
Some numerical values that were used in the simulation study which is reported
next, are shown in Table 1.

\singlespacing
\begin{center}
	
	$%
	\begin{tabular}
	[c]{c|c|c|c|}\cline{2-4}
	& $\kappa=1$ & $\kappa=2$ & $\kappa=3$\\\hline
	\multicolumn{1}{|c|}{$S_{2}$} & 0.90 & 0.60 & 0.46\\\hline
	\multicolumn{1}{|c|}{$S_{1}$} & 0.81 & 0.36 & 0.21\\\hline
	\multicolumn{1}{|c|}{$S_{1/2}$} & 0.65 & 0.13 & 0.04\\\hline
	\multicolumn{1}{|c|}{$t_{3}$} & 1.07 & 0.64 & 0.46\\\hline
	\multicolumn{1}{|c|}{$t_{2}$} & 1.00 & 0.55 & 0.38\\\hline
	\end{tabular}
	\ \ \ $
	
	$%
	\begin{array}
	[c]{cl}%
	\text{Table 1:} & \text{Scale parameter }\sigma\text{ solving
		(\ref{kappa to stable sig}) and (\ref{kappa to student sig})}%
	\end{array}
	$
\end{center}

\doublespacing

\subsection{Symmetric distributions}

\label{Symmetric distributions}

We used standard normal control limits in $50,000$ Monte Carlo realizations of
the two-sided CUSUM in each of five underlying symmetric unimodal
distributions: wrapped Student $t$-distributions with $2$ and $3$ degrees of
freedom and three wrapped stable distributions with indexes $\alpha=2$ (the
wrapped normal distribution), $\alpha=1$ (the wrapped Cauchy distribution,
which is also the wrapped Student $t$-distribution with $1$ degree of freedom)
and $\alpha=1/2$ (the wrapped symmetrized L\'{e}vy distribution). Except for
the wrapped normal, these are wrapped versions of heavy-tailed symmetric
distributions on the real line. Each of the distributions was standardized to
concentrations of $\kappa=1,\ 2$ and $3$ by specifying the scale parameter
$\sigma$ (see Table 1) in accordance with (\ref{kappa to stable sig}) and
(\ref{kappa to student sig}). Two sets of simulations were run. In the first
set, the CUSUMs were initiated at $n=11$, the first $m=10$ observations
serving to establish initial estimates of the unknown parameters. In the
second set we took $m=25$, initiating the CUSUM at $n=26$.

We present in Tables 2.1 and 2.2 aggregated sets of results representing the
general picture. (Detailed tables are given in the supplementary material to
this paper.) Each entry is the average of five estimated in-control ARLs, one
from each of the five distributions. The number in brackets shows the range of
the five estimates. The tables show the results for reference values $\zeta=0$
and $\zeta=0.25$.

\singlespacing
\begin{center}%

	$%
	\begin{tabular}
	[c]{c|c|c|c|c|c|c|}\cline{2-7}
	& \multicolumn{3}{|c|}{$\zeta=0$} & \multicolumn{3}{|c|}{$\zeta=0.25$}\\\hline
	\multicolumn{1}{|c|}{$ARL_{0}$} & $\kappa=1$ & $\kappa=2$ & $\kappa=3$ &
	$\kappa=1$ & $\kappa=2$ & $\kappa=3$\\\hline
	\multicolumn{1}{|c|}{$250$} & 242 (2) & 243 (5) & 242 (4) & 236$^{\ast}$ (4) &
	233$^{\ast}$ (2) & 225$^{\ast}$ (20)\\\hline
	\multicolumn{1}{|c|}{$500$} & 490 (3) & 491 (6) & 491 (10) & 493 (9) & 483
	(8) & 464$^{\ast}$ (52)\\\hline
	\multicolumn{1}{|c|}{$1000$} & 1037 (9) & 1039 (14) & 1042 (20) & 1018 (7) &
	997 (30) & 958$^{\ast\dagger}$ (117)\\\hline
	\end{tabular}
	\bigskip
	$
		\begin{tabular}
		[c]{cl}%
		Table 2.1: & Average in-control ARL of the non-parametric CUSUM\\
		& in five symmetric distributions ($m=10$). The number in\\
		& brackets is the range of the five estimates.
	\end{tabular}
	\bigskip
	
	$%
	\begin{tabular}
	[c]{c|c|c|c|c|c|c|}\cline{2-7}
	& \multicolumn{3}{|c|}{$\zeta=0$} & \multicolumn{3}{|c|}{$\zeta=0.25$}\\\hline
	\multicolumn{1}{|c|}{$ARL_{0}$} & $\kappa=1$ & $\kappa=2$ & $\kappa=3$ &
	$\kappa=1$ & $\kappa=2$ & $\kappa=3$\\\hline
	\multicolumn{1}{|c|}{$250$} & 244 (2) & 244 (6) & 245 (7) & 242 (4) & 239
	(3) & 234$^{\ast}$ (8)\\\hline
	\multicolumn{1}{|c|}{$500$} & 492 (4) & 493 (7) & 493 (10) & 498 (9) & 491
	(7) & 478 (28)\\\hline
	\multicolumn{1}{|c|}{$1000$} & 1039 (11) & 1041 (10) & 1045 (17) & 1024
	(13) & 1005 (26) & 971 (82)\\\hline
	\end{tabular}
	\ \ \bigskip$
	
		\begin{tabular}
		[c]{cl}%
		Table 2.2: & Average in-control ARL of the non-parametric CUSUM\\
		& in five symmetric distributions ($m=25$). The number in\\
		& brackets is the range of the five estimates.
	\end{tabular}
\end{center}

\doublespacing

All but the four starred estimates shown in the Tables lie within $5\%$ of the
nominal value. The exceptions, which all lie within $10\%$, occur at
$\zeta=0.25$ and predominantly at the smaller warmup $m=10$. In the cell
marked $^{\ast\dagger}$ the five estimates were $874,\ 959,\ 976,\ 988$ and
$991$, the outlier $874$ coming from the very heavy tailed L\'{e}vy
distribution. In fact, all three discrepancies in this column are attributable
to a substantial underestimate from the L\'{e}vy distribution \ Clearly, the
CUSUM is very near distribution free overall when a reference constant close
to zero is used. With a larger reference constant, as the concentration
increases so does the variation in true ARL between distributions. This
behaviour can be explained to a large extent by reference to the martingale
central limit theorem upon which the construction of the CUSUM rests. If the
summand $\xi_{n}$ is replaced by $\xi_{n}\mp\zeta,$ the cumulative sums take
the form $\ S_{k}\mp k\zeta$ where
\begin{equation}
S_{k}=\sum_{n=m+1}^{m+k}
\hat{\xi}_{n},\ k\geq1 \label{partial sums}%
\end{equation}
and $\zeta$ is positive. The rationale behind the construction of the CUSUM
consists essentially in replacing \ the discrete time process $S_{k}/h=%
{\textstyle\sum\nolimits_{n=m+1}^{m+k}}
\hat{\xi}_{n}/h,$ $k\geq1$, where $h$ is the control limit, by a continuous
time Brownian motion process, $W(t),\ t>0$. This is effected by changing the
time scale. We identify $k$ with $th^{2}$ where $h$ is the control limit, and
then replace $S_{k}/h$ by $W(th^{2})/h$, which has the same distribution as
$W(t)$. Similarly, $k\zeta$ is replaced by $th^{2}\zeta/h=th\zeta$. Thus,
$(S_{k}\mp k\zeta)/h,\ k\geq1$, is replaced by $W(t)-th\zeta$. The validity of
this procedure requires that $h$ tends to $\infty$. Now, if $\zeta$ is
positive and $h\rightarrow\infty$ then the drift term $th\zeta\rightarrow
\infty$, which makes the resulting CUSUM useless. To avoid this effect,
$\zeta$ must be chosen to be $O(1/h)$, which in practical terms means that
$\zeta$ should be a small positive number or zero.

Next, the effect of any Phase I estimation on the in-control Phase II
performance of the CUSUM needs to be considered. Given $\hat{\zeta}$, let
$\hat{h}$ be the control limit which gives a standard normal CUSUM an
in-control ARL value $ARL_{0}$. The simulation results in Tables 2 and 3
together with the ensuing discussion indicate that the resulting Phase II
CUSUM is near distribution free provided that the reference constant is
suitably close to zero. Thus, regardless of the form of the underlying
distribution, in such cases the true Phase II in-control ARL will be nearly
constant and acceptably close to the nominal value $ARL_{0}$. This behaviour
is in stark contrast to that of parametric CUSUMs where estimating unknown
parameters from Phase I data and then pretending that the Phase I estimate is
the true value, affects irrevocably the in-control ARL of the Phase II CUSUM.
Then there is no guarantee that the in-control ARL will be equal to, or even
near, the nominal value. This point has been made repeatedly in the published
literature, most recently by Keefe, et al. (2015, Introduction section) and
Saleh et Al. (2016). Hawkins and Olwell (1998, pages 159-160) give a realistic
example in which the true in-control ARL of a normal distribution CUSUM, with
variance estimated from Phase I data, differs by two orders of magnitude from
the nominal value.

In this connection, and to illustrate further the in-control behaviour of the
nonparametric CUSUM, we present next a result that is representative of a
general pattern. Consider a situation in which data arise from a wrapped
$t_{3}$ distribution with concentration parameter $\kappa$ - see
(\ref{kappa to student sig}). CUSUMs with reference constants $\zeta=0$ and
$\zeta=0.25$ and nominal in-control ARL $500$ are run at $\kappa=1$ and
$\kappa=3$. A Phase I sample of size $m=30$ is used in each case to obtain an
initial value $B_{m}^{\ast}$ of the sequence of denominators in the summands
$\hat{\xi}_{n}$ see (\ref{xi modified}). The "true" in-control ARLs, estimated
from $50,000$ Monte Carlo trials in each instance, are shown in Table
3.

\singlespacing
\begin{center}

	$%
	\begin{tabular}
	[c]{c|c|c|}\cline{2-3}
	& $\zeta=0$ & $\zeta=0.25$\\\hline
	\multicolumn{1}{|c|}{$\kappa=1$} & 492 & 499\\\hline
	\multicolumn{1}{|c|}{$\kappa=3$} & 492 & 482\\\hline
	\end{tabular}
	\ \ \ $
	
	\begin{tabular}
		[c]{ll}%
		Table 3: & Estimated in-control ARL of direction CUSUM for data\\
		& from a wrapped $t_{3}$ distribution with concentration parameter $\kappa.$ \\
		& Warmup $m=30$ and based on $50,000$ Monte Carlo trials.
	\end{tabular}
\end{center}
\doublespacing
\bigskip

In each of the six instances the $50,000$ values of $B_{m}^{\ast}$ were
grouped into bins of unit length and the average of the corresponding run
lengths in each bin calculated. Figure 1 shows plots of these average run
lengths against the midpoints of the bins together with confidence intervals
of width equal to three estimated standard errors (Bins containing fewer than
$100$ observations, which contain the less commonly occurring values of
$B_{m}^{\ast}$, are not shown.) The figure thus provides a representation of
the Phase II in-control ARL, \textit{conditional upon} the Phase I estimate
$B_{m}^{\ast}$. It is only at the combination $\kappa=3,\ \zeta=0.25$. that
the Phase II in-control ARL exhibits substantial systematic variation away
from the corresponding unconditional value in Table 3.

\singlespacing
\begin{figure}
\begin{center}
	$%
	\begin{tabular}
	[c]{c|c|c|}\cline{2-3}
	& $\zeta=0$ & $\zeta=0.25$\\\hline
	\multicolumn{1}{|c|}{}& & \\
	\multicolumn{1}{|c|}{$\kappa=1$} &
	
	\includegraphics[trim = 15mm 5mm 13mm 5mm, clip,height=2.1in]{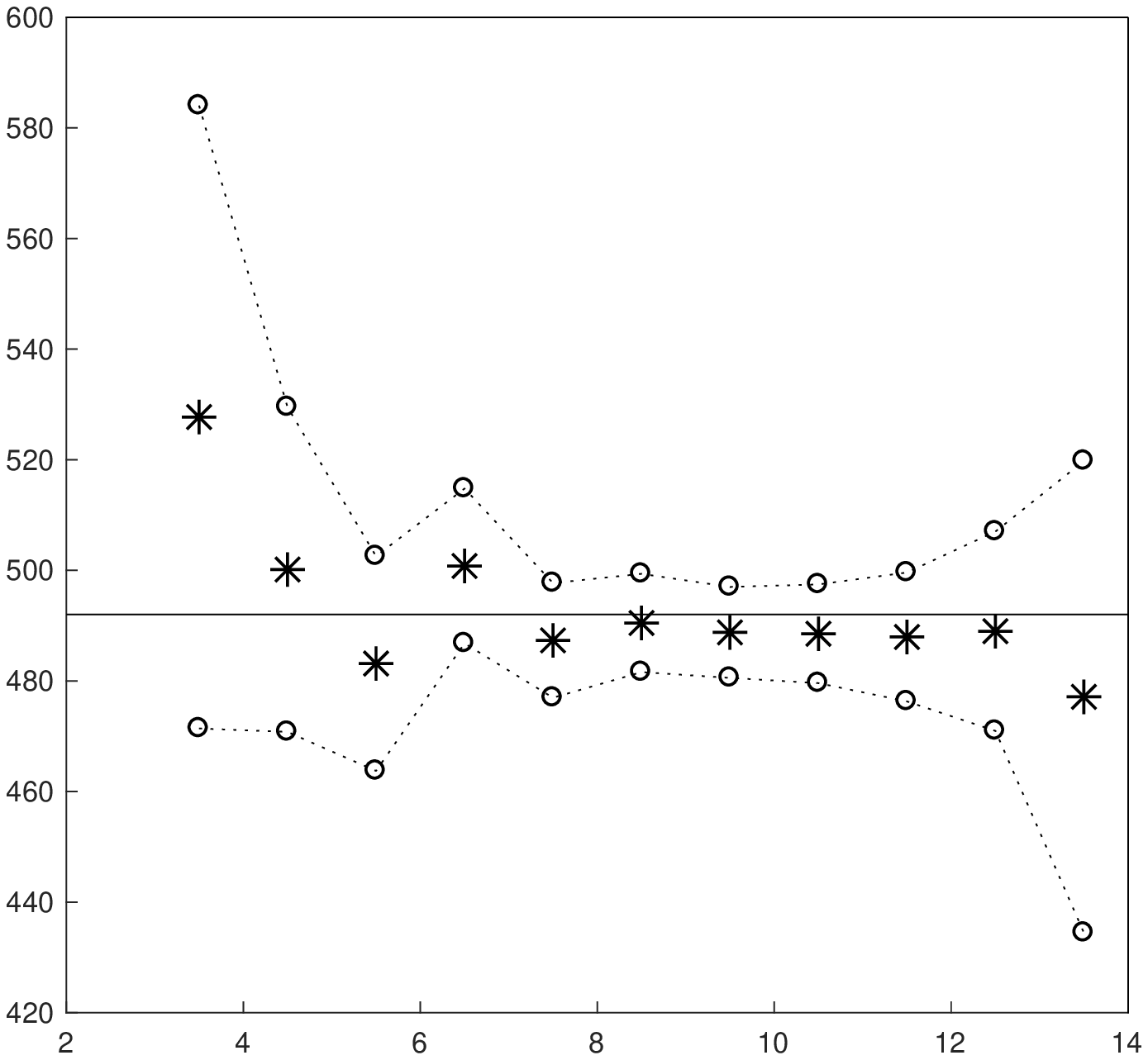}%
	&
	\includegraphics[trim = 15mm 5mm 13mm 5mm, clip,height=2.1in]{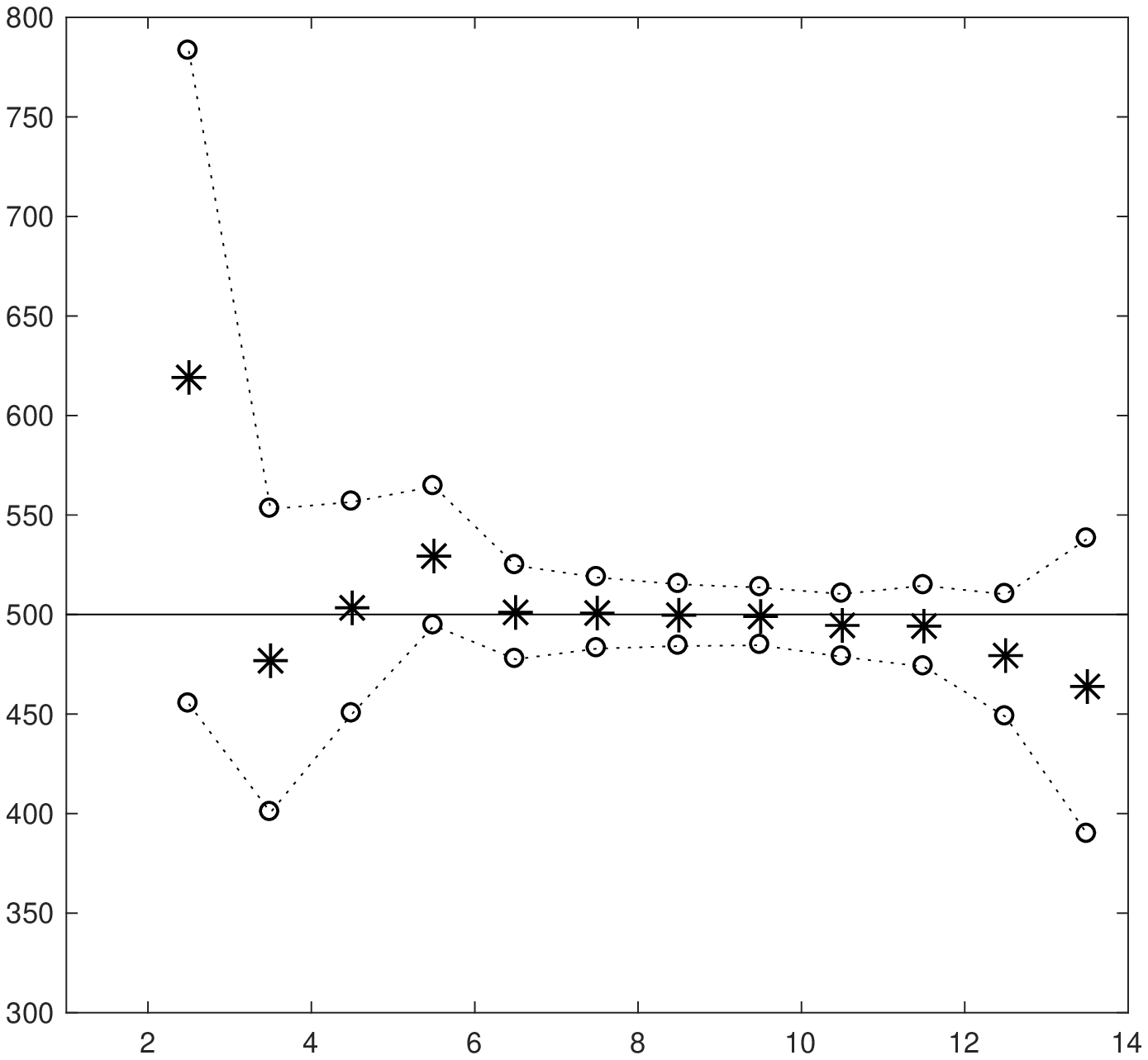}%

	\\\hline
	\multicolumn{1}{|c|}{}& & \\
	\multicolumn{1}{|c|}{$\kappa=3$} &
	\includegraphics[trim = 15mm 5mm 13mm 5mm, clip,height=2.1in]{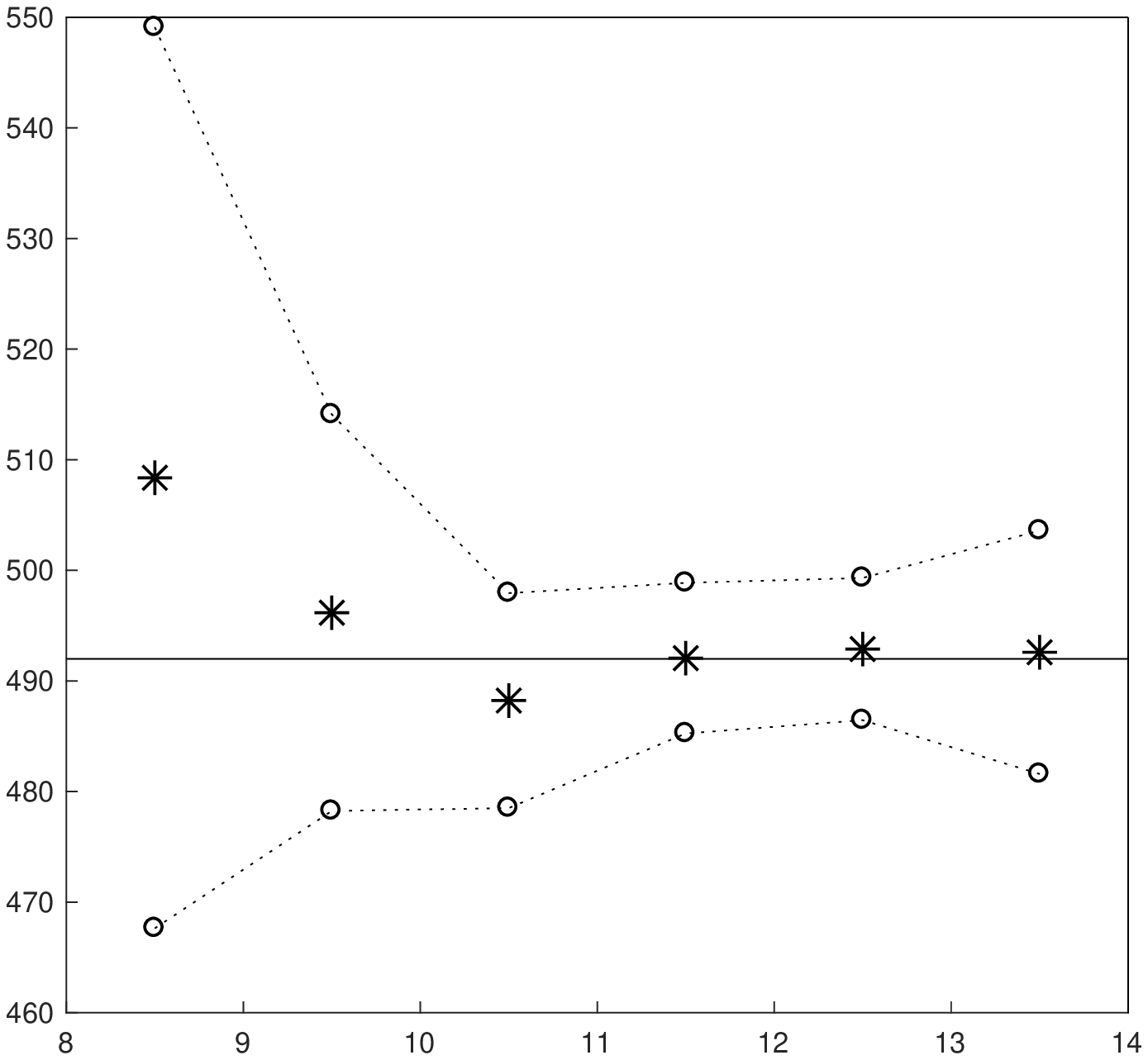}%

	&
	\includegraphics[trim = 15mm 5mm 13mm 5mm, clip,height=2.1in]{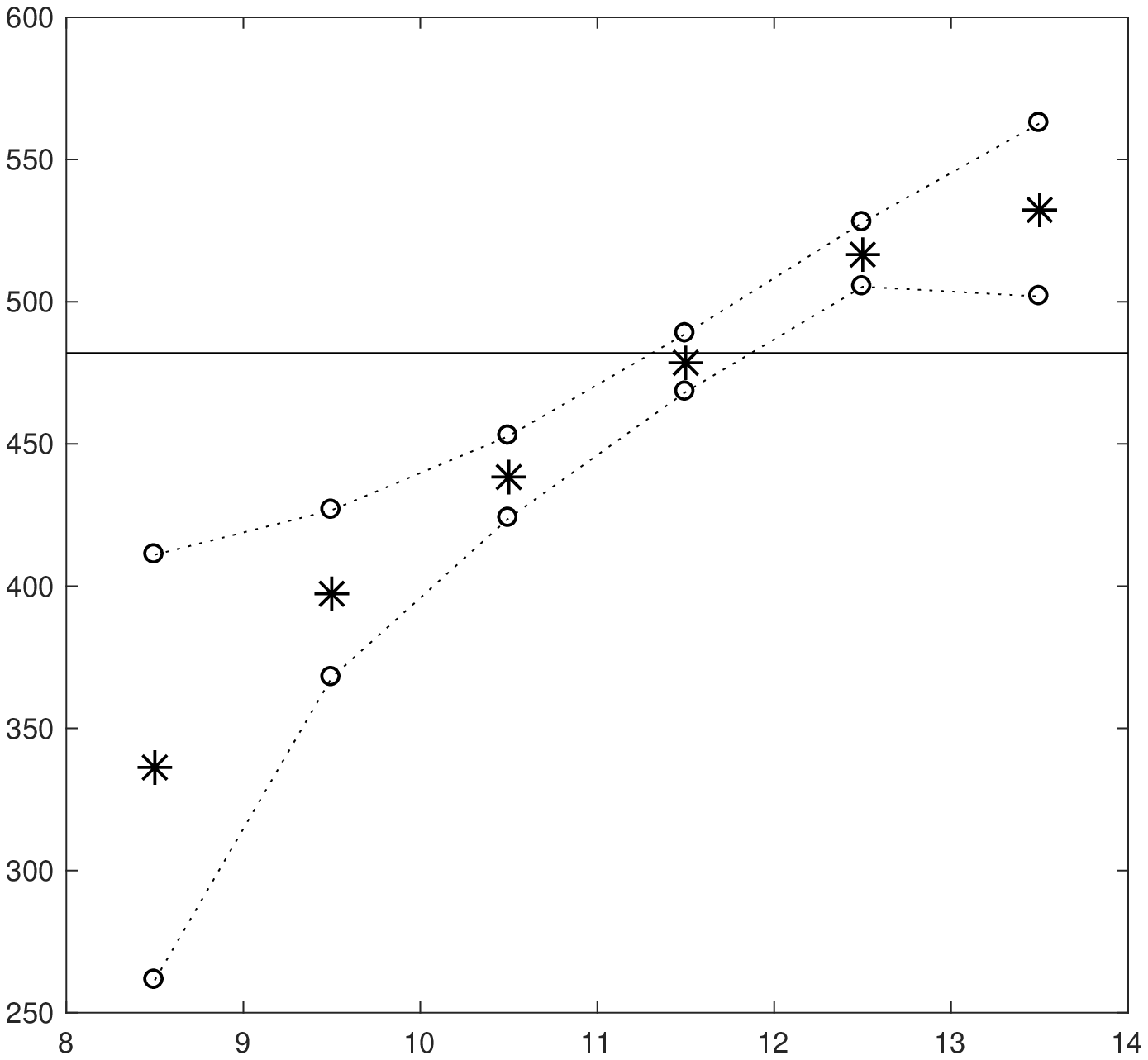}%

	\\\hline
	\end{tabular}
	\ \ \ $
	
	$%
	\begin{array}
	[c]{cl}%
	\\ \text{Figure 1:} & \text{In-control ARL (on the vertical axis), conditional
		upon the value of }B_{30}^{\ast}\text{ }\\
	& \text{(on the horizontal axis), for two concentrations\ }\kappa\text{ and
		two reference }\\
	& \text{values }\zeta\text{ in wrapped }t_{3}\text{ distributions. The stars
		denote the ARL values}\\
	& \text{and the dotted lines are }95\%\text{ confidence intervals}%
	\end{array}
	$
\end{center}

\end{figure}
\doublespacing

\subsection{Asymmetric distributions}

\label{Asymmetric distributions}

To assess the effect of skewness in the underlying distribution on the
in-control ARL, we generated data from wrapped skew-normal distributions
(Pewsey, 2000) with mean direction zero and skewness parameters $\lambda=2$
(lightly skewed), $\lambda=7$ (moderately skewed) and $\lambda=\infty$
(heavily skewed), wrapped skew-stable Cauchy- and L\'{e}vy distributions with
skewness parameters $\beta=0.75$ and $1.0$ (Jammallamadaka and SenGupta, 2001,
Section 2.2.8) and from wrapping skew-$t$ distributions (Jones and Faddy,
2003) with $2$ and $3$ degrees of freedom and skewness parameters
$\lambda=2,\ 7$ and $\infty\,$. The aggregated results are in Tables 4.1 and
4.2. Comparing the results with those in Tables 2.1 and 2.2, we see that the
general pattern is the same. The main contributors to the apparent degradation
seen at $\zeta=0.25$,\ $\kappa=3$ are the excessively skewed distributions,
namely the wrapped skew-normal and $t$-distributions with skewness parameter
$\lambda=\infty$ and the wrapped L\'{e}vy distribution with skewness parameter
$\beta=1$. These distributions produce estimates that are consistently
substantially lower than the rest. This is perhaps not too surprising if one
takes account of their shape. The supplementary material to this paper has a
Figure showing a plot of a wrapped skew-t density with $2$ degrees of freedom
and skewness parameters $\lambda=0,\ 2\ $and $7$ at $\kappa=3$. The extreme
skewness and high concentration at $\lambda=7$ magnifies the deleterious
effect that a large reference value has on the approximation to the nominal
in-control ARL (Section \ref{Symmetric distributions}, first paragraph after
Table 2.2). The degradation noted above largely disappears when such highly
skewed distributions are eliminated from consideration.

\singlespacing
\begin{center}
	
	\begin{tabular}
	[c]{c|c|c|c|c|c|c|}\cline{2-7}
	& \multicolumn{3}{|c|}{$\zeta=0$} & \multicolumn{3}{|c|}{$\zeta=0.25$}\\\hline
	\multicolumn{1}{|c|}{$ARL_{0}$} & $\kappa=1$ & $\kappa=2$ & $\kappa=3$ &
	$\kappa=1$ & $\kappa=2$ & $\kappa=3$\\\hline
	\multicolumn{1}{|c|}{$250$} & 241 (2) & 240 (4) & 238 (7) & 235 (5) & 228
	(11) & 217 (25)\\\hline
	\multicolumn{1}{|c|}{$500$} & 489 (4) & 487 (6) & 484 (9) & 490 (8) & 474
	(29) & 448 (71)\\\hline
	\multicolumn{1}{|c|}{$1000$} & 1039 (11) & 1036 (9) & 1031 (13) & 1013 (13) &
	979 (61) & 915 (178)\\\hline
	\end{tabular}
	\ \ \ \ \bigskip%
	
		\begin{tabular}
		[c]{cl}%
		Table 4.1: & Average in-control ARL of the non-parametric CUSUM\\
		& in thirteen asymmetric distributions ($m=10$). The number\\
		& in brackets is the range of the thirteen estimates.
	\end{tabular}
	
	$%
	\begin{tabular}
	[c]{c|c|c|c|c|c|c|}\cline{2-7}
	& \multicolumn{3}{|c|}{$\zeta=0$} & \multicolumn{3}{|c|}{$\zeta=0.25$}\\\hline
	\multicolumn{1}{|c|}{$ARL_{0}$} & $\kappa=1$ & $\kappa=2$ & $\kappa=3$ &
	$\kappa=1$ & $\kappa=2$ & $\kappa=3$\\\hline
	\multicolumn{1}{|c|}{$250$} & 243 (2) & 242 (3) & 242 (4) & 240 (3) & 235
	(7) & 229 (22)\\\hline
	\multicolumn{1}{|c|}{$500$} & 491 (5) & 490 (6) & 489 (7) & 494 (7) & 484
	(25) & 463 (59)\\\hline
	\multicolumn{1}{|c|}{$1000$} & 1039 (13) & 1038 (10) & 1038 (11) & 1019
	(17) & 988 (65) & 935 (161)\\\hline
	\end{tabular}
	$
	
\bigskip

	\begin{tabular}
	[c]{cl}%
	Table 4.2: & Average in-control ARL of the non-parametric CUSUM\\
	& in thirteen asymmetric distributions ($m=25$). The number\\
	& in brackets is the range of the thirteen estimates.
\end{tabular}

\end{center}

\doublespacing

\subsection{Choice of reference constant}

\label{Choice of ref const}

We saw in Sections \ref{Symmetric distributions} and
\ref{Asymmetric distributions} that the CUSUM exhibits good in-and
out-of-control behaviour throughout when a small positive reference constant
$\zeta$ is used. In analogy with a normal distribution CUSUM, one would expect
the CUSUM to then be quite adept at detecting small changes but less effective
if the change is of substantial magnitude. In the latter case, efficient
detection of a change requires use of a larger reference constant. Again in
analogy with a normal distribution CUSUM, an appropriate choice of reference
constant for efficient detection of a rotation of size $\geq\delta_{0}$
\textit{could} be%
\[
\zeta=\frac{\mathrm{E}[\sin(X+\delta_{0}-\nu)-\sin(X-\nu)]}{\sqrt{\mathrm{Var}%
		[\sin(X-\nu)]}},
\]
which can be estimated from some in-control Phase I data $X_{1},\ldots,X_{m}$
by%
\begin{equation}
\hat{\zeta}=\frac{\delta_{0}}{2}\times\frac{m^{-1}\sum\nolimits_{j=1}%
	^{m}\sin(X_{j}+\delta_{0}-\hat{\nu}_{m})}{\sqrt{m^{-1}\sum\nolimits_{j=1}%
		^{m}\sin^{2}(X_{j}-\hat{\nu}_{m})}}. \label{ref_constant}%
\end{equation}
Clearly, the variability of the estimator $\hat{\zeta}$ will depend on both
the size $m$ of the in-control Phase I sample and on the type of the unknown
underlying distribution. If $\hat{\zeta}$ turns out to be too large given the
known limitations of the CUSUM, one could use a reference value $\hat{\zeta}$
$\leq0.25$, say, and solve for $\delta_{0}$ from (\ref{ref_constant}). This
$\delta_{0}$ would serve as an indication of the magnitude of change that the
CUSUM could be expected to detect efficiently.

\subsection{Out-of-control properties}

\label{Direction OOCARL properties}

While the in-control behaviour of the CUSUM is similar to that of a CUSUM for
normal data on the real line, the same is not true in respect of its
out-of-control behaviour. In fact, we show next that a consequence of the
continual updating of the mean direction estimator $\hat{\nu}_{n}$ from
(\ref{mean direction}) is that after a change of mean direction the CUSUM will
return eventually to what appears to be an in-control state. This behaviour is
similar to that of self-starting CUSUMs for linear data, and is a warning to
users of the need for corrective action as soon as a change is diagnosed- see
Hawkins and Olwell (1998, Section 7.1).

Suppose there is a rotation of size $\delta$ from $n=\tau+1$ onwards and set
$Y_{i}=X_{i+\tau}+\delta,\ i\geq1$ \ Then, using the approximations%
\[
\frac{1}{\tau+k}\approx0\ and\ \frac{k}{\tau+k}\approx1
\]
for large $k$ and fixed $\tau\geq m$, the mean direction estimated from the
data $X_{1},\ldots,X_{\tau},Y_{1},\ldots,Y_{k}$ is%
\begin{align*}
\hat{\nu}_{\tau+k}  &  =\text{atan2}\left(  \frac{S_{\tau}+\sum_{i=1}%
	^{k}\sin Y_{i}}{\tau+k},\frac{C_{\tau}+\sum_{i=1}^{k}\cos Y_{i}%
}{\tau+k}\right) \\
& \\
&  \approx\text{atan2}\left(  \frac{\sum_{i=1}^{k}\sin Y_{i}}%
{k},\frac{\sum_{i=1}^{k}\cos Y_{i}}{k}\right)  :=\hat{\nu}_{k}(Y),
\end{align*}
which is the estimated mean direction of $Y_{i},\ 1\leq i\leq k$. Thus, for
sufficiently large $k$, $\hat{\nu}_{\tau+k}$ is in effect estimating the mean
direction of the post-change observations $Y_{1},\ldots,Y_{k}$. Consequently,%
\[
\hat{\xi}_{\tau+k+1}\approx\frac{\sin(Y_{k+1}-\hat{\nu}_{k}(Y))}{\sqrt
	{k^{-1}\sum_{i=1}^{k}\sin^{2}(Y_{i}-\hat{\nu}_{k}(Y))}}%
\]
which, because of its rotation invariance, has the same distribution as the
in-control variable $\hat{\xi}_{k}$.

A further consequence of this behaviour is that, in the absence of a
substantial amount of in-control Phase I data there is no simple manner in
which to assess, a priori, the out-of-control ARL
\[
\mathrm{E}[N-\tau|N>\tau]
\]
of the CUSUM. Here $N-\tau$ is the time taken for an alarm to be raised after
a change has occurred, the expected value being calculated upon an assumption
of no false alarms prior to the change. Nevertheless, simulation results
indicate that the out-of-control ARL of the two-sided CUSUM behaves in an
appropriate manner, namely that the out-of-control ARL is less than the
in-control $ARL_{0}$ and that it decreases as the size of the shift increases
from $0$ to $\pi/2$. For shifts of size in excess of $\pi/2,$the ARL starts
increasing again. This behaviour is a result of the periodic nature of the
CUSUM summand. Furthermore, that choosing $\zeta=0$ leads to substantially
larger out-of-control ARLs compared to those produced by small positive
reference constants.

To illustrate that the general pattern of out--of-control ARL behaviour mimics
that of a normal distribution CUSUM, Table 5 gives out--of-control ARL
estimates from $10,000$ simulations involving in each case shifts $\delta$ of
sizes ranging from $\pi/8$, to $7\pi/8$ in a wrapped Cauchy distribution with
$\kappa=2$, a warmup sample size $m=25$ and reference constants $\zeta=0$,
$\zeta=0.125$ and $\zeta=0.25$. The in-control ARL was $1,000$ throughout. The
results are for shifts induced respectively at observation $\tau=100$ and at
observation $\tau=200$.\bigskip

\singlespacing
\begin{center}

	\begin{tabular}
	[c]{c|c|c|c|c|c|c|}\cline{2-7}
	& \multicolumn{3}{|c}{$\tau=100$} & \multicolumn{3}{|c|}{$\tau=200$%
	}\\\cline{2-7}
	& $\zeta=0$ & $\zeta=0.125$ & $\zeta=0.25$ & $\zeta=0$ & $\zeta=0.125$ &
	$\zeta=0.25$\\\hline
	\multicolumn{1}{|c|}{$\delta=\pi/8$} & $123$ & $49$ & $82$ & $82$ & $37$ &
	$39$\\\hline
	\multicolumn{1}{|c|}{$\delta=\pi/4$} & $50$ & $17$ & $14$ & $40$ & $17$ &
	$13$\\\hline
	\multicolumn{1}{|c|}{$\delta=\pi/2$} & $31$ & $11$ & $8$ & $28$ & $11$ &
	$8$\\\hline
	\multicolumn{1}{|c|}{$\delta=3\pi/4$} & $38$ & $16$ & $12$ & $37$ & $15$ &
	$12$\\\hline
	\multicolumn{1}{|c|}{$\delta=7\pi/8$} & $58$ & $29$ & $26$ & $61$ & $31$ &
	$28$\\\hline
	\end{tabular}
	\ \ 
	
		\begin{tabular}
		[c]{ll}%
		\\ Table 5: & Estimated out--of-control ARL of direction CUSUM for data\\
		& from a wrapped Cauchy distribution with concentration\\
		& parameter $\kappa=2.$ Warmup $m=25.$ Changepoints $\tau=100$\\
		& and $\tau=200$.%
	\end{tabular}
\end{center}
\doublespacing

If a sufficiently large amount of in-control Phase I data are available to
allow a non-trivial nonparametric estimate of the underlying density to be
made (Taylor, 2008), the in-control and out-of-control properties of the CUSUM
can be fathomed by sampling from the estimated density.

\subsection{Bimodal distributions}

\label{Biimodal}

Thus far attention has focussed on unimodal distributions. However, many of
the properties of the proposed CUSUM remain intact when the underlying
distribution is multimodal. Here, we restrict attention to bimodal densities
of the form
\begin{equation}
f(\theta)=pg(\theta)+(1-p)g(\theta-\mu_{0}) \label{Bimodal model}%
\end{equation}
with $1/2\leq p<1$ and a unimodal density $g$ on the circle. Since the
concentration of $f$ will be less than that of $g$, one finds that the
approximation to the nominal in-control ARL often improves markedly, even at a
reference constant $0.25$. For instance, let $g$ in (\ref{Bimodal model}) be a
von Mises density with high concentration $\kappa=3.42$ and mean $0$. Then, if
$p=1$ (which is the unimodal case), and with $\zeta=0.25$ and a nominal
in-control ARL of $500$, the estimated true in-control ARL is $461$. On the
other hand if $p=1/3$ and $\mu_{0}=-3\pi/4$, in which case $f$ is bimodal with
concentration equal to $1$, the estimated true in-control ARL of $492$ is much
closer to the nominal value.

On the other hand, the ability of the CUSUM to detect a change of size
$\delta\neq0$ decreases as $\mu_{0}$ in (\ref{Bimodal model}) nears $\pm\pi$
and vanishes when $f$ in (\ref{Bimodal model}) is antipodal, that is, when
$p=1/2$ and $|\mu_{0}|=\pi$. Put another way, the CUSUM is then unable to
distinguish between $f(\theta)$ and $f(\theta-\delta)$. The ostensible reason
for this behaviour is that an antipodal distribution does not possess a well
defined mean or median \ Nevertheless, a non-trivial CUSUM will result upon
replacing the data $X_{i}$ by $2X_{i}$. This replacement transforms
$f(\theta)$ to $g(\theta/2)/2$, which is unimodal - see, for instance,
Jammalamadaka and SenGupta (2001, page 48).

\section{Concentration change}

\label{Concentration change}

For data $X_{1},\ldots,X_{n}$ from a von Mises($\nu,\kappa$) distribution,
locally most powerful tests of the hypothesis $\kappa=\kappa_{0}\ (\neq0)$ are
based on the statistic $\sum_{i=1}^{n}\cos(X_{i}-\nu)$. However, the fact that $\kappa$ is not a scale parameter of
the distribution of $X$ complicates matters. Hawkins and Lombard (2017) showed
that even if the mean direction $\nu$ is known, control limits for a specified
in-control ARL in a von Mises CUSUM for detecting change away from $\kappa
_{0}$ depend upon $\kappa_{0}$. Nonetheless, the locally most powerful test
statistic suggests application of a CUSUM based on
\[
V_{n}^{\prime}=\cos(X_{n}-\hat{\nu}_{n-1}),\ n\geq1.
\]
Again, there are purely non-parametric interpretations of $V_{n}^{\prime}$,
devoid of any reference to a von Mises distribution. For instance, since%
\[
V_{n}^{\prime}=(C_{n-1}/R_{n-1})\cos X_{n}+(S_{n-1}/R_{n-1})\sin X_{n},
\]
we see that $V_{n}^{\prime}$ is the (signed) length of the projection of the
vector $y_{n}=(\sin X_{n},\cos X_{n})$ in the direction $\hat{\nu}%
_{n-1}\approx\nu$ of the unit vector\newline$(S_{n-1}/R_{n-1},C_{n-1}%
/R_{n-1})$. If the concentration increases (decreases) after $n=\tau$, the
average of $V_{\tau+1}^{\prime},\ldots,V_{\tau+k}^{\prime}$ will tend to be
greater (smaller) than the average of $V_{1}^{\prime},\ldots,V_{\tau}^{\prime
}$. Another non-parametric interpretation rests on the fact that $R_{n}^{2}$
in (\ref{R^2}) is a frequently used non-parametric measure of concentration in
a sample $X_{1},\ldots,X_{n}$. Simple algebra shows that the relative change
in $R_{n-1}^{2}$ brought about by the next observation $X_{n}$ is%
\[
\frac{R_{n}^{2}}{R_{n-1}^{2}}-1=\frac{2V_{n}^{\prime}}{R_{n-1}}+\frac
{1}{R_{n-1}^{2}}_{,}%
\]
again justifying consideration of $V_{n}^{\prime}$.

Proceeding in much the same manner as in Section \ref{Construction}, a CUSUM
of%
\begin{equation}
\hat{\xi}_{n}^{\prime}=\frac{\cos(X_{n}-\hat{\nu}_{n-1})-R_{n-1}%
	/(n-1)}{B_{n-1}^{\prime}} \label{xi disp}%
\end{equation}
where%
\[
B_{n}^{\prime}=\sqrt{n^{-1} \sum\nolimits_{i=1}^{n} \cos^{2}(X_{i}-\hat{\nu}_{n})-R_{n}^{2}/n^{2}},
\]
is suggested to detect a change in concentration.

A change in the numerical value of $\kappa$ has a much greater effect on the
denominator $B_{n-1}^{\prime}$ in (\ref{xi disp}) than a change of direction
has on the denominator $B_{n-1}$ in (\ref{xi direction cusum}). Furthermore,
the distribution of $V_{n}^{\prime}$ is heavily skewed. Consequently, a CUSUM
based on $\hat{\xi}_{n}^{\prime}$ cannot be expected to have a near
distribution free in-control ARL over a wide range of reference values.
Indeed, simulation results indicate that one is essentially restricted to
$\zeta=0\ $and a large ($\geq500$) nominal in-control ARL if a satisfactory
degree of in-control distribution freeness is to be had over the families of
distributions considered in Section \ref{In control properties}.

\section{Applications}

\label{Examples}

In the two applications treated here we define the sample mean direction of
data $X_{1},\ldots,X_{n}$ by
\[
\hat{\nu}_{n}=\mathrm{atan2}\left(\sum\nolimits_{i=1}^{n}
\sin X_{i},\sum\nolimits_{i=1}^{n} \cos X_{i}\right)
\]
and the sample concentration, by%
\[
\hat{\kappa}_{n}=A^{-1}\left(n^{-1} \sum\nolimits_{i=1}^{n}
\cos(X_{i}-\hat{\nu}_{n})\right)=A^{-1}\left(  \frac{R_{n}}{n}\right) ,%
\]
in analogy with (\ref{kappa defn}). After a CUSUM signals, we estimate the
changepoint $\tau$ in the conventional manner. That is, if the CUSUM signals
with $D^{+}\ $($D^{-}$) at $n=N$, the changepoint estimate is the last index
$n$ $<N$ at which $D_{n}^{+}$ $=0$ ($D_{n}^{-}=0$). Both data sets are
included in the supplementary material to the paper.

\subsection{Acrophase data}

The data, kindly provided by Dr. Germaine Cornelissen of the University of
Minnesota Chronobiology Laboratory, come from ambulatory monitoring equipment
worn by a patient suffering from episodes of clinical depression. The time at
which systolic blood pressure reaches its maximum value on a given day is
called the acrophase. Monitoring the acrophase can provide an automated early
warning of a possible medical condition before it becomes clinically obvious.
We show the results of a two-sided CUSUM analysis with reference constant
$\zeta=0.25$ (recommended reference value from (\ref{ref_constant}) to enable
detection of a $30$ degree, i.e. $\pi/6=0.52$ radian, rotation) and control
limits $h=\pm$ $8.59$, which leads to an in-control ARL of approximately
$500$. The first $m=30$ observations are used to find initial estimates of the
required parameters.

The left-hand panel in Figure 2 shows the CUSUM. The upper CUSUM $D^{+}$
signals at $n=66$ and the changepoint estimate is $\hat{\tau}=57$, that is,
$27$ observations after the warmup period. The right-hand panel in Figure 2
shows the CUSUM after restarting at $n=88$, observations $58$ through $87$
serving as a warmup to estimate the new direction. A sustained decrease in the
lower CUSUM $D^{-}$ is evident. The CUSUM signals at $n=120$, a changepoint
being indicated at $n=110$. Continuing in this manner produces the results in
Table 6, which shows the progress of the CUSUMs as the data accrue. The
estimate of the mean direction and concentration in each segment is shown in
the third and fourth columns of the table.

\singlespacing
\begin{center}
	  $%
	  \begin{array}
	  [c]{cc}%
	
	\includegraphics[height=2.1in]{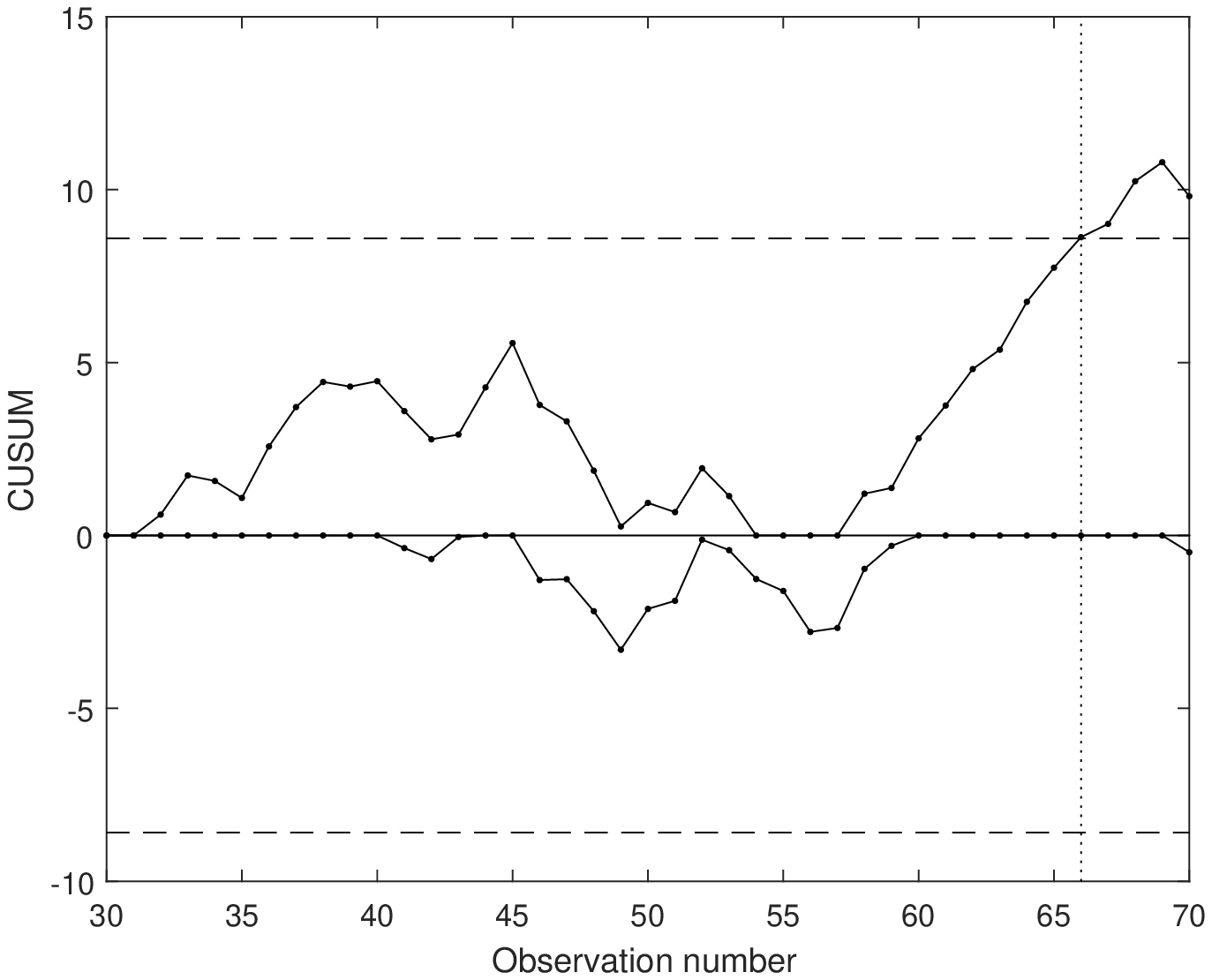}
	&
	
	\includegraphics[height=2.1in]{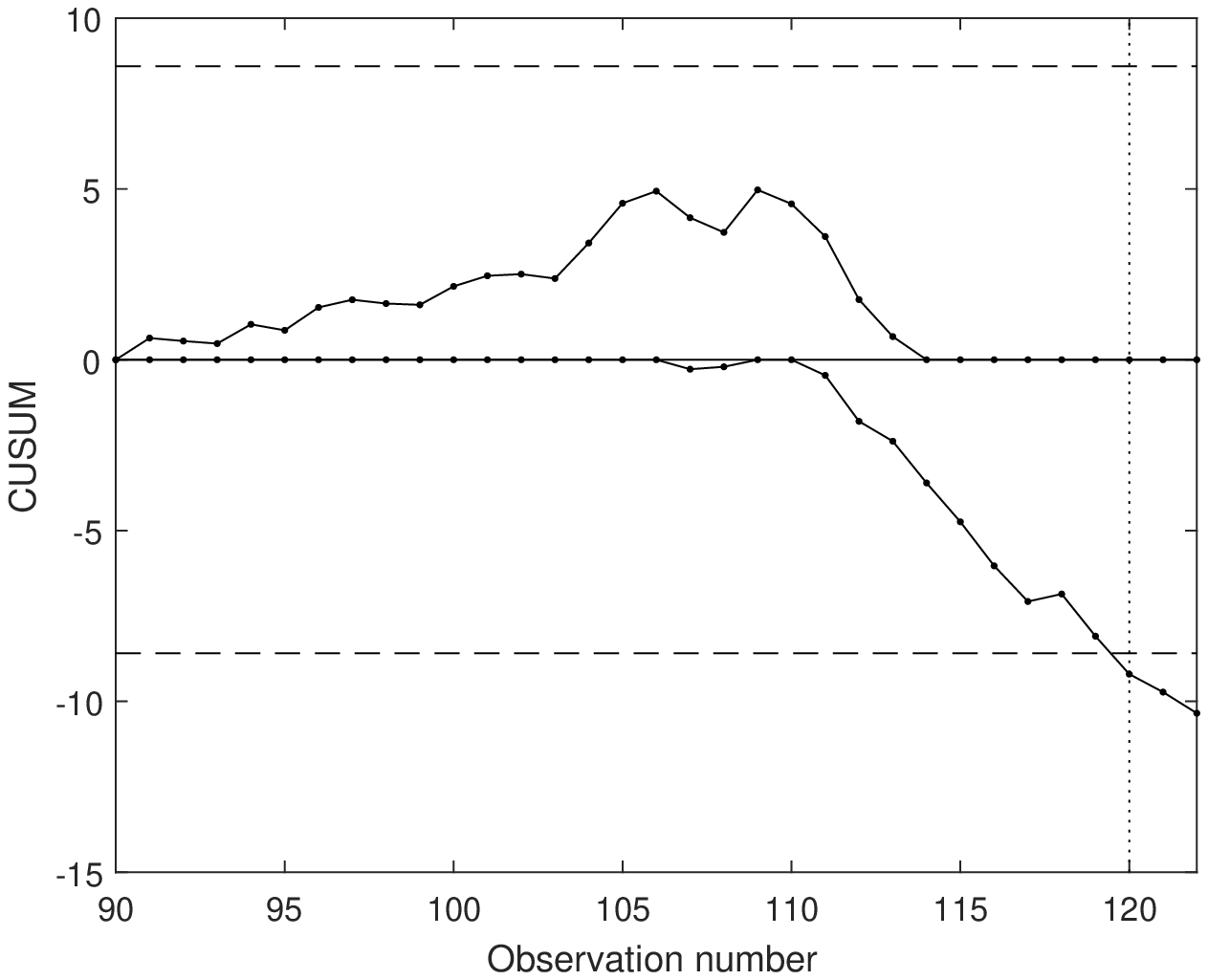}%
	\end{array}
	$

	$%
	\begin{array}
	[c]{cl}%
	\text{Figure 2:} & \text{Direction CUSUMs of acrophase data. Left-hand
		panel:}\\
	& \text{CUSUM after start at }n=31\text{. Right-hand panel: CUSUM}\\
	& \text{after restart at }n=88\text{. The vertical dotted lines indicate}\\
	& \text{the location of the estimated changepoints. The dashed}\\
	& \text{horizontal lines indicate the control limits.}%
	\end{array}
	$%

	\bigskip
	
	\begin{tabular}
		[c]{|c|c|c|c|}\hline
		segment & signal at & $\hat{\nu}$ & $\hat{\kappa}$\\\hline
		$1-57$ & $66$ & $-1.70$ ($263^\circ$) & $1.86$\\\hline
		$58-110$ & $120$ & $-0.76$ ($317^\circ$) & $0.78$\\\hline
		$111-140$ & $178$ & $-1.90$ ($251^\circ$)& $2.60$\\\hline
		$141-241$ & $255$ & $-1.19$ ($292^\circ$) & $2.51$\\\hline
		$242-282$ & $299$ & $-0.90$ ($308^\circ$)& $0.31$\\\hline
		$283-306$ & none & $-.007$ ($360^\circ$)& $1.68$\\\hline
	\end{tabular}
\bigskip

 	\begin{tabular}
 	[c]{cl}%
 	Table 6: & Acrophase data: Progression of CUSUMs
 	\end{tabular}
		
\end{center}
\doublespacing

Figure 3 shows dot plots, constructed after the fact, of the data in the six
identified segments together with an indication of the mean in each segment. A
noticeable feature in this plot is the first two increases followed by a
sudden large decrease to more or less the original mean value. This is
indicative of an external intervention in the treatment of the patient to
reset the acrophase. After that, there follows a sustained increase, this time
without any apparent external intervention. The figure also reveals some
variation between the concentrations within the six segments - see the fourth
column in Table 6. This does not affect the validity of the CUSUM since there
is no assumption that the concentrations in the various segments must all be
the same. In retrospect, it seems that the CUSUM has done a good job of
identifying location changes.

\singlespacing
\begin{figure}[h!]

\begin{center}%
	\begin{tabular}{cc}
		\includegraphics[trim = 15mm 15mm 15mm 5mm, clip,width=65mm]{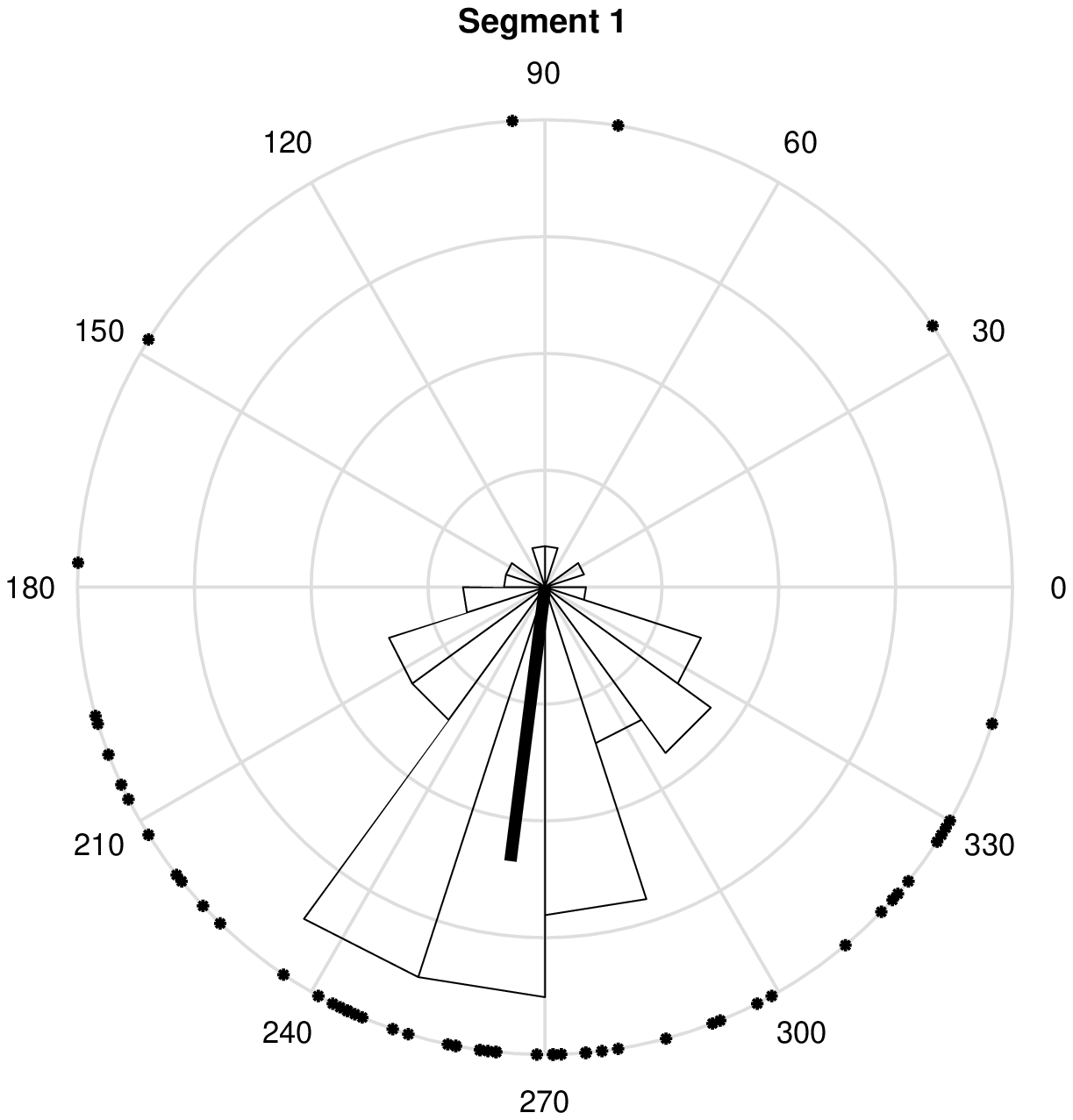} &
		\includegraphics[trim = 15mm 15mm 15mm 5mm, clip,width=65mm]{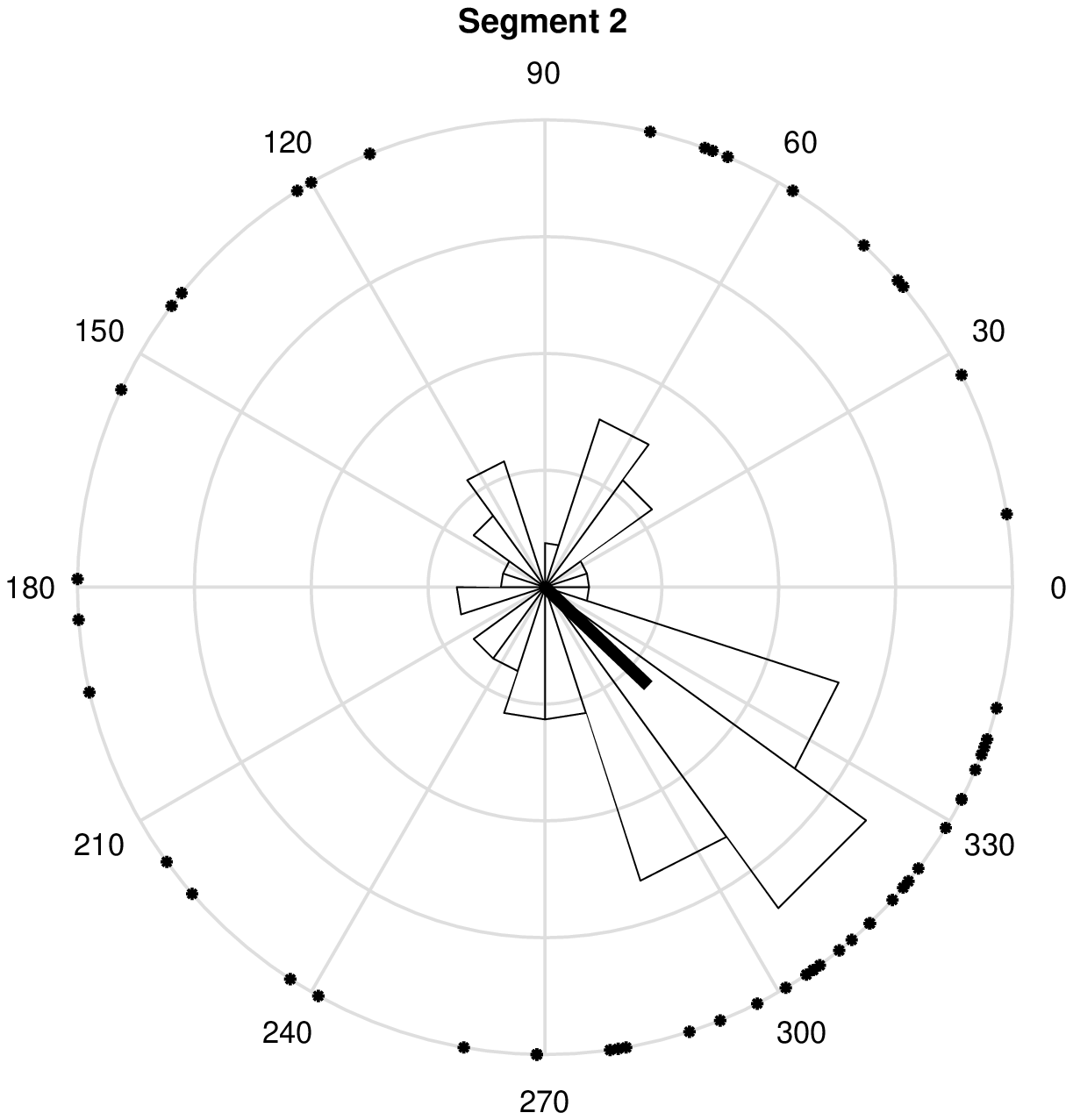} \\
		\includegraphics[trim = 15mm 15mm 15mm 5mm, clip,width=65mm]{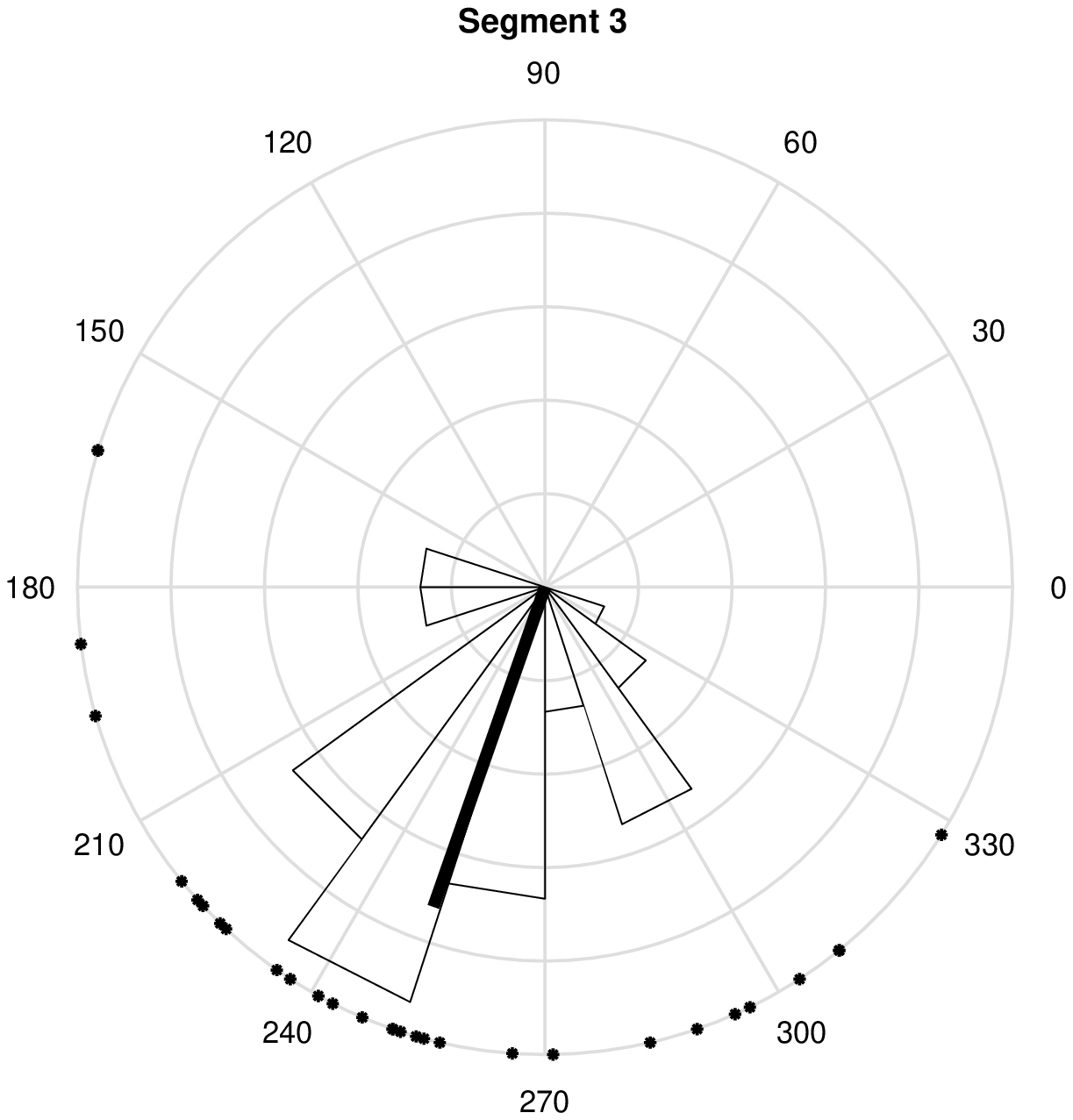} &
		\includegraphics[trim = 15mm 15mm 15mm 5mm, clip,width=65mm]{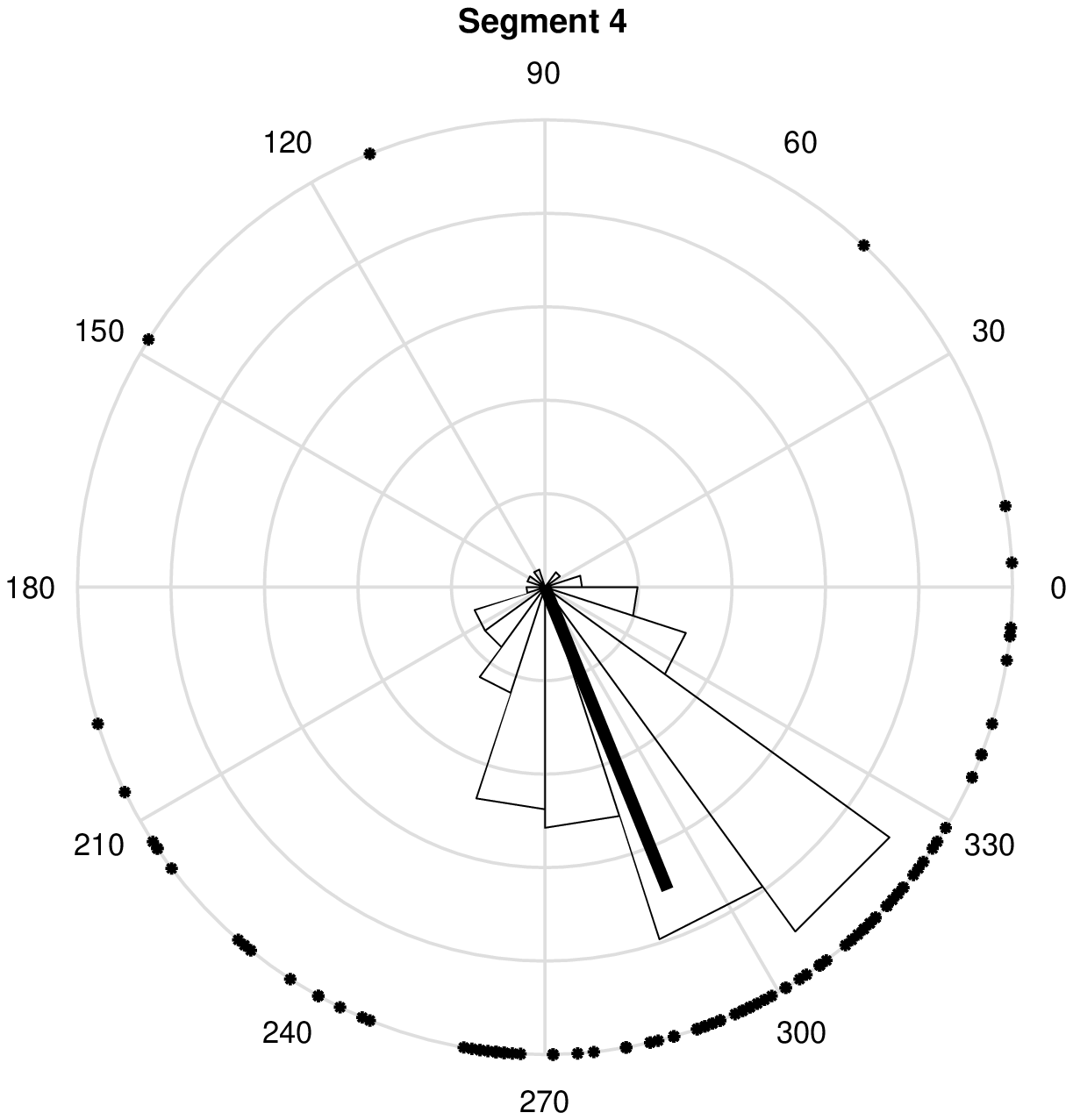} \\
		\includegraphics[trim = 15mm 15mm 15mm 5mm, clip,width=65mm]{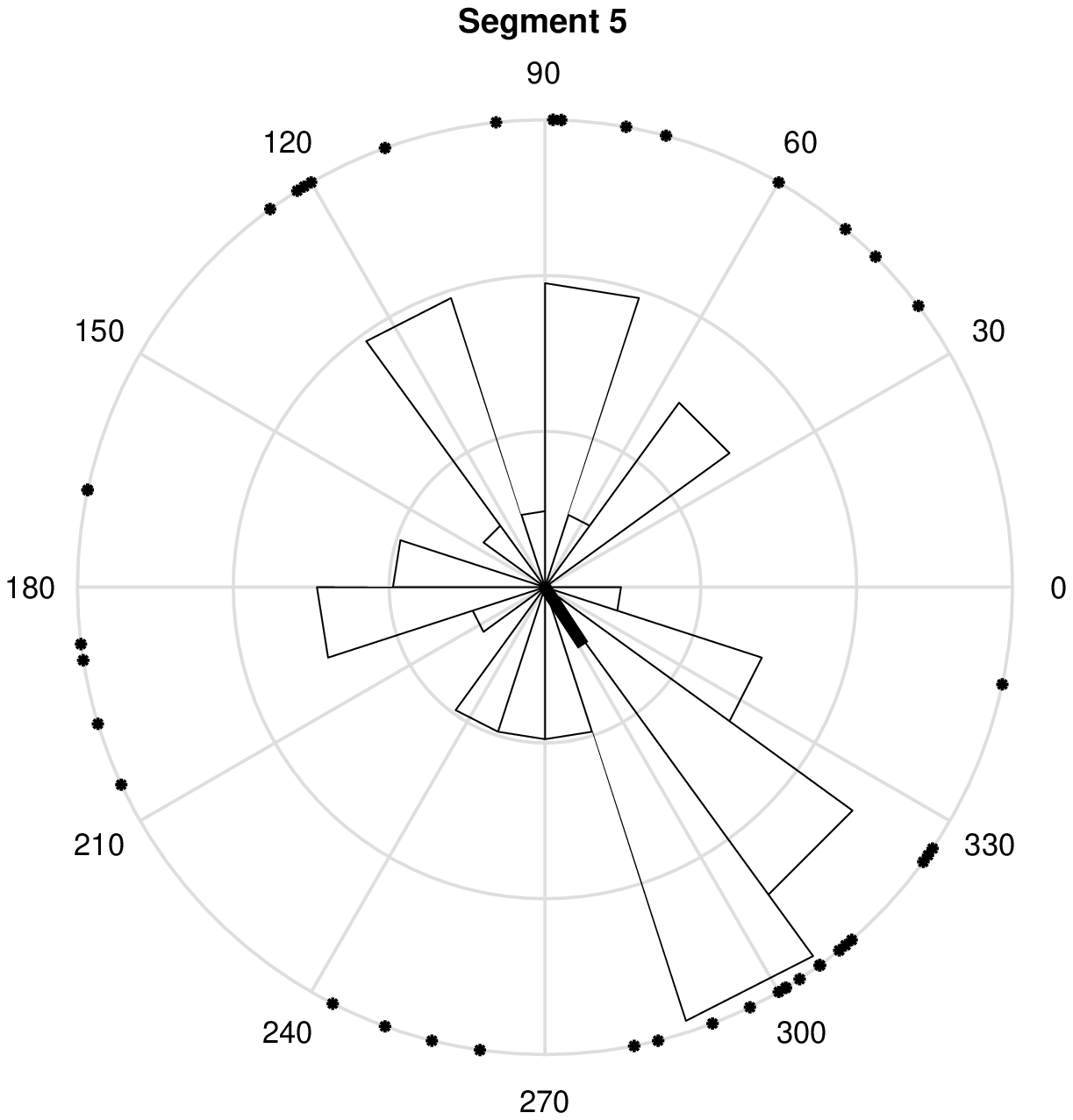} &
		\includegraphics[trim = 15mm 15mm 15mm 5mm, clip,width=65mm]{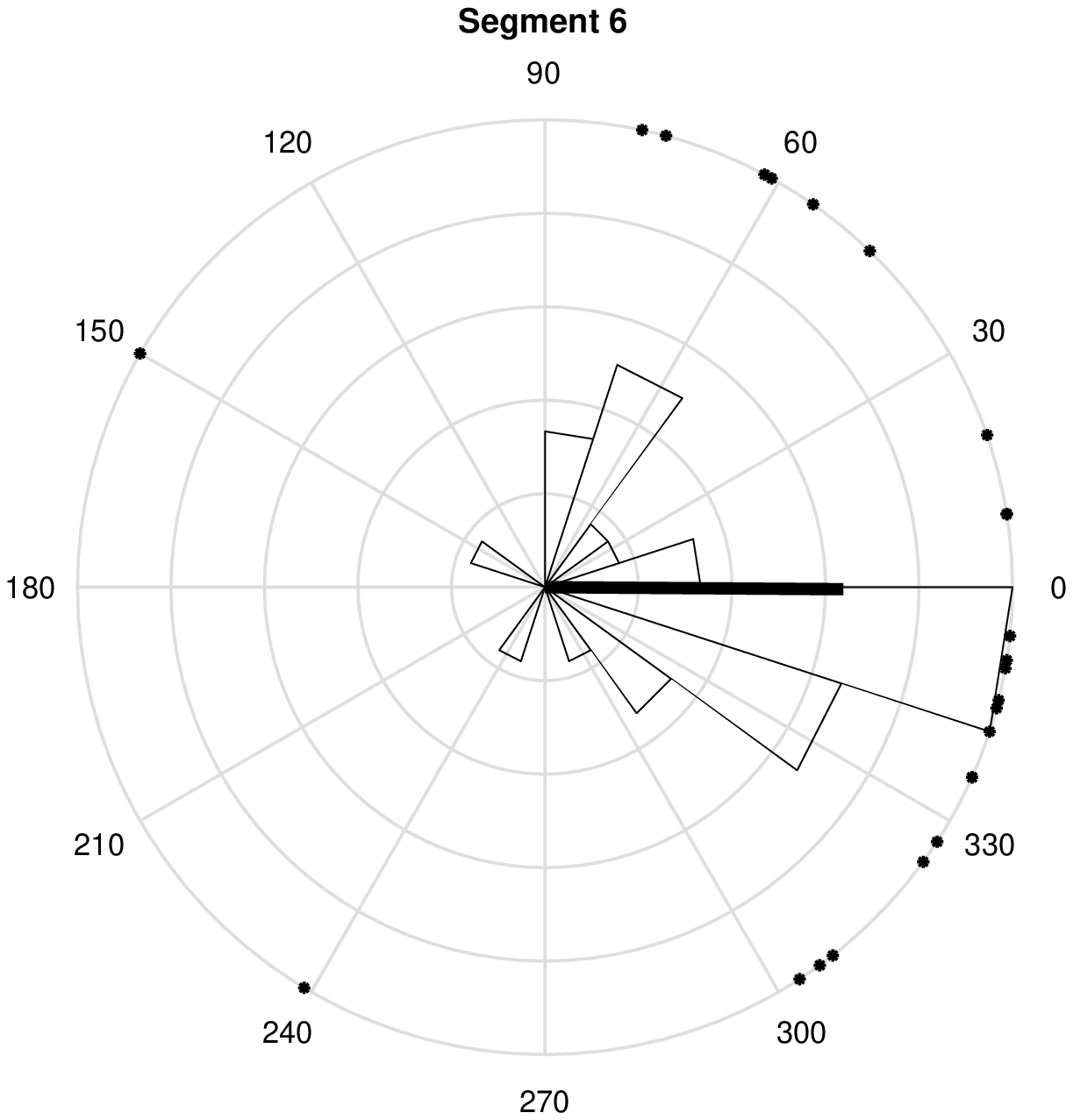} \\
	\end{tabular}

	$%
	\begin{array}
	[c]{cl}%
	\text{Figure 3:} &
	\text{Rose plots of the data in each of the six identified segments of the
acrophase data.}\\
	\end{array}
	$
\end{center}
\end{figure}
\doublespacing

\subsection{Pulsar data}

Lombard and Maxwell (2012) developed a rotation invariant cusum to detect
deviation from a uniform distribution on the circle and applied it to some
data consisting of arrival times of cosmic rays from the vicinity of a pulsar.
The objective is to detect periods of sustained high energy radiation.
Following a standard procedure in Astrophysics, the data were wrapped around a
circle of circumference equal to the period of the pulsar. If no high energy
radiation is present the wrapped data should be more or less uniformly
distributed on the circumference of the circle, while a non-uniform
distribution should manifest itself during periods of high energy radiation.
They found that the first $190$ observations could reasonably be assumed to
have arisen from a uniform distribution. We now apply to observations $191$
through $1250$ the concentration CUSUMs from Section
\ref{Concentration change} of the present paper to detect further changes in
concentration. The in-control ARL of the chart is set at $500$ observations
with reference value $\zeta=0$ (again, the recommended reference value from
(\ref{ref_constant}) to enable detection of a $30$ degree, i.e. $\pi/6=0.52$
radian, rotation) and control limits $\pm30.46$. The first $m=50$ observations
are used to obtain initial estimates of the required means, variances and
covariance of $\sin\ X$ and $\cos\ X$.

The full extent of the concentration CUSUM, without restarts, is shown in
Figure 4. The first signal is at $n=191+495=686$ and the changepoint is
estimated at $n=191+331=522$. The estimated concentration in the segment
$[192,522]$ is $0.35$. Thereafter, the lower CUSUM $D^{-}$ shows a sustained
decrease to the end of the data series. In fact, if the CUSUM is restarted at
$n=523$, a changepoint is indicated at $n=523$. Such a pattern is indicative
of a more or less continuous decrease in concentration as the series
progresses. The estimated concentration of the observations in the segment
$[523,1250]$ is $0.06$, suggesting a uniform distribution in this segment.
Hawkins and Lombard (2015) applied a retrospective segmentation method to
these data. Except for a short segment $[191-207]$, which falls within the
warmup set used to initiate the CUSUM, the results of the CUSUM analysis agree
quite well with their results. The numerical details are shown in Table 7.

\singlespacing
\begin{center}
	\includegraphics[trim = 15mm 5mm 15mm 5mm, clip,width=3.832in]%
	{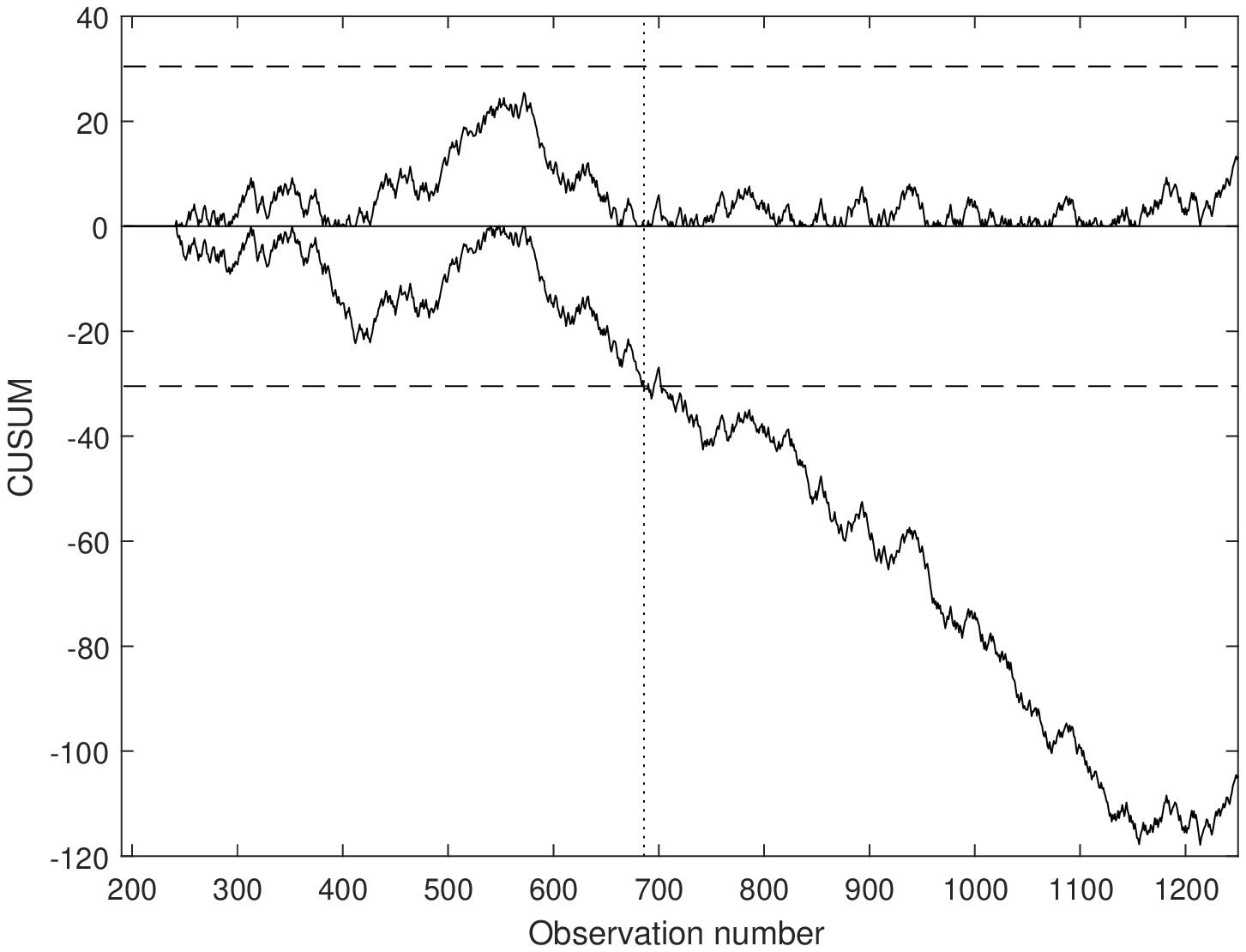}%
\end{center}

\begin{center}
	$%
	\begin{array}
	[c]{cl}%
	\text{Figure 4:} & \text{Concentration CUSUM of the pulsar data.}%
	\end{array}
	\bigskip$%
	
	\begin{tabular}
		[c]{|c|c|c|}\hline
		\multicolumn{3}{|c|}{Retrospective}\\\hline
		segment & $\hat{\nu}$ & $\hat{\kappa}$\\\hline
		191-207 & -0.41 & 1.89\\\hline
		208-573 & -1.58 & 0.35\\\hline
		574-1250 & - & 0.0\\\hline
	\end{tabular}%
	\begin{tabular}
		[c]{|c|c|c|}\hline
		\multicolumn{3}{|c|}{CUSUM}\\\hline
		segment & $\hat{\nu}$ & $\hat{\kappa}$\\\hline
		&  & \\\hline
		191-522 & -1.44 & 0.35\\\hline
		523-1250 & - & 0.06\\\hline
	\end{tabular}

	\begin{tabular}
	[c]{cl}
	\\ Table 7: & Pulsar data. Segments delineated by sequential\\
	& CUSUM and retrospective segmentation
\end{tabular}

\end{center}
\doublespacing

\section{Summary}

\label{Summary}

We develop non-parametric rotation invariant CUSUMs for detecting changes in
the mean direction and concentration of a circular distribution. The CUSUMs
are designed for situations in which the initial mean direction and
concentration are unspecified, the objective being to detect a change from the
initial values, whatever the latter may be. Monte Carlo simulation results
indicate that the CUSUMs have in-control average run lengths that are
acceptably close to the nominal values over a wide class of symmetric and
asymmetric circular distributions. Two applications of the methodology to data
from Health Science and Astrophysics are discussed.\bigskip

\section*{Supplementary Material}Supplementary material for this publication is available on GitHub at \\ \href{https://github.com/cpotgieter/nonparametric-cusums}{https://github.com/cpotgieter/nonparametric-cusums}. The supplementary files consist of a pdf document with detailed simulation results, an Excel file with the datasets used in this paper, and the Matlab code for implementing the CUSUM procedures proposed here.

\section*{Acknowledgement }The authors thank two referees for some valuable
comments that led to an improved presentation of the results in the paper.

\singlespacing


\begin{thebibliography}{99}                                                                                               %
	
	
	\bibitem {Abe Pewsey}Abe, T. and \ Pewsey, A., (2011). Sine-skewed circular
	distributions. \textit{Statistics Papers}, \textbf{52}, 683--707.
	
	\bibitem {Azzalini}Azzalini, A. and Capitanio,A., (2003). Distributions
	generated by perturbation of symmetry with emphasis on a multivariate skew t
	distribution\textit{ Journal of the Royal Statistical Society},\textbf{B},
	\textbf{65}, 367-389.
	
	\bibitem {Fisher}Fisher, N., (1993). Statistical Analysis of Circular data.
	Cambridge: Cambridge University Press.
	
	\bibitem {H and O}Hawkins, D. M., and Olwell, D. H., (1998).
	\textit{Cumulative Sum Charts and Charting for Quality Improvement}. Springer
	Verlag, New York.
	
	\bibitem {HOW}Hawkins, D.M., Olwell, D. H. and Wang, B., (2016). http://cran.r-project.org/web/packages/CUSUMdesign/CUSUMdesign.pdf
	
	\bibitem {HL1}Hawkins, D M. and Lombard, F., (2015). Segmentation of circular
	data, \textit{Journal of Applied Statistics}, \textbf{42} (1), 88-97.
	
	\bibitem {HL2}Hawkins, D M. and Lombard, F., (2017). CUSUM control for data
	following the von Mises distribution. \textit{Journal of Applied Statistics}.
	\textbf{44}:8,
	
	\bibitem {Helland}Helland, I., (1982). Central Limit Theorems for Martingales
	with Discrete or Continuous Time. \textit{Scandinavian Journal of Statistics},
	\textbf{9}, 79-94.
	
	\bibitem {J and SenGupta}Jammalamadaka, S. R., and SenGupta, A., (2001).
	\textit{Topics in Circular Statistics}. Singapore: World Scientific Publishing Company.
	
	\bibitem {Jones Faddy}Jones, M.C. and Faddy, M.J., (2003). A Skew Extension of
	the t-Distribution, with Applications. \textit{Journal of the Royal
		Statistical Society}, \textbf{B}, \textbf{65},159-174.
	
	\bibitem {Keefe et al}Keefe, M.J., Woodall, W.H. and Jones-Farmer, L.A.,
	(2015). The Conditional In-Control Performance of Self-Starting Control
	Charts. \textit{Quality Engineering}, \textbf{27}, 488-499.
	
	\bibitem {Knoth}Knoth, S., (2017). spc:
	Statistical Process Control --
	Collection of Some Useful
	Functions. R, package version
	0.5.4.
	https://CRAN.R-project.org/package=spc.
	
	\bibitem {L and M}Lombard, F. and Maxwell, R.K., (2012). A CUSUM Procedure to
	Detect Deviations from Uniformity in Angular Data. \textit{Journal of Applied
		Statistics}, \textbf{39}, 1871-1880.
	
	\bibitem {LHP}Lombard, F., Hawkins, D.M. and Potgieter, C.J. (2018).
	Sequential rank CUSUM charts for angular data. \textit{Computational
		Statistics and Data Analysis}, \textbf{105}, 268-279.
	
	\bibitem {Mardia & Jupp}Mardia, K.V. and Jupp, P.E., (2000). Directional
	Statistics. Chichester: John Wiley and Sons.
	
	\bibitem {Matlab}Mathworks: Matlab Version 2016b.
	
	\bibitem {Nolan}Nolan, J.P, (2015). \textit{Stable Distributions-Models for
		Heavy Tailed Data}. Boston: Birkhauser. Note: In progress, Chapter 1 online at
	academic.2.american.edu/\symbol{126}jpnolan.
	
	\bibitem {Page}Page, E.S., (1954). Continuous Inspection Schemes.
	\textit{Biometrika}, \textbf{41}, 100-115.
	
	\bibitem {Pewsey}Pewsey, A., (2000). The wrapped skew-normal distribution on
	the circle. \textit{Communications in Statistics - Theory and Methods},
	\textbf{29} (11), 2459-2472.
	
	\bibitem {Saleh et Al}Saleh, N.A., Zwetsloot, I.M., Mahmood, A.M. and Woodall,
	W.H. (2016). CUSUM charts with controlled conditional performance under
	estimated parameters. \textit{Quality Engineering}, \textbf{28}, 402-425.
	
	\bibitem {Taylor}Taylor, C.C., (2008). Automatic Bandwidth Selection for
	Circular Density Estimation. \textit{Computational Statistics and Data
		Analysis}, \textbf{52}, 3493-3500.
	
	\bibitem {U and J}Umbach, D. and Jammalamadaka, S.R., (2011). Building
	asymmetry into circular distributions. \textit{Statistics and Probability
		Letters}, \textbf{79}, 659-663.
\end{thebibliography}
\end{document}